\pdfoutput=1

\IfFileExists{./prepreamble-amspreprint.sty}{\RequirePackage[packages,theorems,changes]{./prepreamble-amspreprint}}{}

\makeatletter
\IfFileExists{./scoop-latex/scoop-packages.sty}{\providecommand*{\input@path}{}\edef\input@path{{./scoop-latex/}\input@path}}{}
\@namedef{ver@minted.sty}{}
\@namedef{opt@minted.sty}{}
\makeatother
\PassOptionsToPackage{style = numeric-comp, backend = biber}{biblatex}
\PassOptionsToPackage{commentmarkup = footnote}{changes}
\PassOptionsToPackage{inline}{enumitem}
\RequirePackage{algpseudocode}

\documentclass[english,lineno]{amsart}

\RequirePackage{biblatex}
\RequirePackage{ltxcmds}
\IfFileExists{./preamble-amspreprint.sty}{\RequirePackage[packages,theorems,changes]{./preamble-amspreprint}}{}

\usepackage{scoop-local}

\IfFileExists{./postpreamble-amspreprint.sty}{\RequirePackage[packages,theorems,changes]{./postpreamble-amspreprint}}{}

\makeatletter
\@ifpackageloaded{changes}{
\definechangesauthor[name = {Karina Koval}, color = {red!80!black}]{KK}
\definechangesauthor[name = {Roland Herzog}, color = {blue!80!black}]{RH}
\definechangesauthor[name = {Robert Scheichl}, color = {orange!80!black}]{RS}
}{}
\makeatother

\makeatletter
\@ifpackageloaded{hyperref}{%
	\hypersetup{
		pdftitle = {Tractable Optimal Experimental Design using Transport Maps},
		pdfauthor = {Karina Koval, Roland Herzog, Robert Scheichl},
		pdfkeywords = {optimal experimental design, Bayesian inverse problems, uncertainty quantification, transportation of measures, low-rank tensor decomposition, tensor trains},
		pdfcreator = {Created using the Scoop Template Engine version 1.2.0.}
	}
}{
	\pdfinfo{
		/Title (Tractable Optimal Experimental Design using Transport Maps)
		/Author (Karina Koval, Roland Herzog, Robert Scheichl)
		/Subject ()
		/Keywords (optimal experimental design, Bayesian inverse problems, uncertainty quantification, transportation of measures, low-rank tensor decomposition, tensor trains)
		/Creator (Created using the Scoop Template Engine version 1.2.0.)
	}
}
\makeatother

\title[Tractable Optimal Experimental Design using Transport Maps]{Tractable Optimal Experimental Design using Transport Maps}

\author{Karina Koval\orcidlink{0000-0001-9741-4530}}
\address[K. Koval]{Interdisciplinary Center for Scientific Computing, Heidelberg University, 69120 Heidelberg, Germany}
\email{karina.koval@iwr.uni-heidelberg.de}
\urladdr{https://scoop.iwr.uni-heidelberg.de}

\author{Roland Herzog\orcidlink{0000-0003-2164-6575}}
\address[R. Herzog]{Interdisciplinary Center for Scientific Computing and Institute for Mathematics, Heidelberg University, 69120 Heidelberg, Germany}
\email{roland.herzog@iwr.uni-heidelberg.de}
\urladdr{https://scoop.iwr.uni-heidelberg.de}

\author{Robert Scheichl\orcidlink{0000-0001-8493-4393}}
\address[R. Scheichl]{Institute for Mathematics and Interdisciplinary Center for Scientific Computing, Heidelberg University, 69120 Heidelberg, Germany}
\email{robert.scheichl@uni-heidelberg.de}
\urladdr{https://katana.iwr.uni-heidelberg.de/people/rob/}

\thanks{This work has been partially funded by Carl Zeiss-Stiftung through the project \enquote{Model-Based AI: Physical Models and Deep Learning for Imaging and Cancer Treatment}. RS and KK would like to thank the Isaac Newton Institute for Mathematical Sciences for support and hospitality during the programme on \emph{Future Data-Driven Engineering} when part of the work on this paper was undertaken, supported by EPSRC grant number EP/R014604/1.}

\date{}

\dedicatory{}

\begin{document}

\begin{abstract}
We present a flexible method for computing Bayesian optimal experimental designs (BOEDs) for inverse problems with intractable posteriors.
The approach is applicable to a wide range of BOED problems and can accommodate various optimality criteria, prior distributions and noise models. 
The key to our approach is the construction of a transport-map-based surrogate to the joint probability law of the design, observational and inference random variables. 
This order-preserving transport map is constructed using tensor trains and can be used to efficiently sample from (and evaluate approximate densities of) conditional distributions that are required in the evaluation of many commonly-used optimality criteria. 
The algorithm is also extended to sequential data acquisition problems, where experiments can be performed in sequence to update the state of knowledge about the unknown parameters. 
The sequential BOED problem is made computationally feasible by preconditioning the approximation of the joint density at the current stage using transport maps constructed at previous stages. 
The flexibility of our approach in finding optimal designs is illustrated with some numerical examples inspired by disease modeling and the reconstruction of subsurface structures in aquifers.

\end{abstract}


\makeatletter
\ltx@ifpackageloaded{hyperref}{%
\subjclass[2010]{\href{https://mathscinet.ams.org/msc/msc2020.html?t=62K05}{62K05}, \href{https://mathscinet.ams.org/msc/msc2020.html?t=62F15}{62F15}, \href{https://mathscinet.ams.org/msc/msc2020.html?t=65K10}{65K10}, \href{https://mathscinet.ams.org/msc/msc2020.html?t=65L09}{65L09}, \href{https://mathscinet.ams.org/msc/msc2020.html?t=65N21}{65N21}, \href{https://mathscinet.ams.org/msc/msc2020.html?t=15A69}{15A69}}
}{%
}
\makeatother

\maketitle

\section{Introduction}
\label{section:introduction}

The Bayesian approach to parameter estimation is widespread throughout the sciences and engineering.
In the Bayesian framework, a-priori knowledge (encoded in a prior measure) of some unknown\breakslash unobservable parameters is updated using experimental data and a mathematical model.
Hence, the solution to the Bayesian inverse problem is a posterior probability law describing the updated state of knowledge conditioned on the observed data.
The quality of the solution is highly dependent on the observed experimental data --- a poorly-chosen experimental setup can lead to uninformed posteriors not straying far from the prior, whereas well-chosen experimental designs can lead to well-informed posteriors characterized by high degrees of certainty.

In many Bayesian inverse problems stemming from real-world phenomena, there are limitations on the number of experiments that can be performed or the amount of data that can be acquired.
These limitations could originate from physical or monetary constraints.
For example, tsunami warning systems rely on pressure sensors tethered to the ocean floor near trenches, and groundwater monitoring systems require drilling wells deep into the ground --- both scenarios are characterized by expensive means of data acquisition.
In such settings, it is crucial to allocate the limited resources well, choosing experimental conditions that maximize the \enquote{quality} of the measured data.

Optimal experimental design (OED) provides a rigorous mathematical framework for addressing the question of how to design experimental conditions for optimal parameter inference.
Standard OED references in the point estimation setting include~\cite{Pazman:1986:1, Ucinski:2005:1, Pukelsheim:2006:1, AtkinsonDonevTobias:2007:1}, and the techniques found therein have gained much traction within the last few decades in the Bayesian inverse problems community.
In OED, experimental designs are chosen to optimize some utility function that assesses the amount of information that could be gleamed from performing any feasible experiment.
The utility functions are user-specified and problem-dependent, though there is a plethora of commonly used criteria.
For any design, these commonly used criteria (\eg, \nobreakseq{A-,} \nobreakseq{D-,} E-optimality~\cite{Alexanderian:2021:1}) evaluate the statistical quality of the resulting posterior distribution.
The overarching goal is to choose designs or experimental conditions that minimize the level of uncertainty, or maximize information, in the posterior.

Solving the OED optimization problem is challenging for Bayesian inverse problems governed by models that involve systems of ordinary differential equations (ODEs) or even partial differential equations (PDEs).
Since the Bayesian inverse problem is merely a sub-problem of OED, all the numerical challenges present in Bayesian inference due to, \eg, the large number of inference parameters or the high cost in evaluating the PDE/ODE models, are inherited.
Many efficient and scalable algorithms for solving these challenging high-dimensional (or infinite-dimensional) OED problems have been developed in recent years~\cite{HuanMarzouk:2013:1, AlexanderianPetraStadlerGhattas:2014:1, HuanMarzouk:2014:1, AlexanderianGloorGhattas:2016:1, AlexanderianPetraStadlerGhattas:2016:1, NeitzelPieperVexlerWalter:2019:1, FosterJankowiakOMearaTehRainforth:2020:1, Alexanderian:2021:1, AlexanderianPetraStadlerSunseri:2021:1, FosterIvanovaMalikRainforth:2021:1, AlexanderianNicholsonPetra:2022:1, AttiaConstantinescu:2022:1, AttiaLeyfferMunson:2022:1, WuOLearyRoseberryChenGhattas:2022:1}.
A large portion of these techniques (see, \eg,~\cite{HaberHoreshTenorio:2008:1, AlexanderianPetraStadlerGhattas:2014:1, AlexanderianSaibaba:2018:1, AttiaAlexanderianSaibaba:2018:1, NeitzelPieperVexlerWalter:2019:1, KovalAlexanderianStadler:2020:1, AlexanderianPetraStadlerSunseri:2021:1, AttiaConstantinescu:2022:1}) are formulated for \emph{linear} Bayesian inverse problems with additive Gaussian noise models and Gaussian priors.
For such problems, the parameter enters linearly into the parameter-to-observable (PTO) map and the posteriors are Gaussian, hence simple to characterize.

In this article, we focus on the design of experimental conditions for Bayesian inverse problems governed by \emph{nonlinear} PTO maps resulting in Bayesian inverse problems characterized by non-Gaussian posteriors.
The nonlinear parameter-to-observable map introduces some unique challenges that are not present in the linear OED setting, even under the assumption of an additive Gaussian noise model --- the most notable of which are lack of a closed-form expression for the utility function, and the dependence of the optimality criterion on the observed data.
The latter issue is often circumvented by finding designs that work well on average for all realizations of the data, \ie, by optimizing the expected value of the chosen utility function.
However, even after this simplification, the challenge of approximating the resulting objective function remains.
A few common approaches to alleviate this involve: linearization techniques (\eg, using a Laplace approximation as in~\cite{LongScavinoTemponeWang:2013:1,AlexanderianPetraStadlerGhattas:2016:1,WuChenGhattas:2023:1}) or employing purely sample-based Monte Carlo estimators (\eg, nested Monte Carlo approximation of the expected information gain, which can be made computationally feasible via surrogates of the parameter-to-observable map~\cite{HuanMarzouk:2013:1, WuOLearyRoseberryChenGhattas:2022:1}).\label{page:challenges}
Methods based on the Laplace approximation have been developed for various utility functions.
They offer scalability for large-dimensional problems and are suitable for weakly nonlinear problems.
However, they do in general require gradients (and Hessians) of the log-posterior density, as well as maximum a-posteriori estimators for a large number of data samples, though approaches that avoid numerical optimization at the expense of additional error are available (see, \eg~\cite[Section~4.4]{WuChenGhattas:2023:1}).
On the other hand, nested Monte Carlo estimators to the expected information gain (EIG) are asymptotically unbiased and consistent~\cite{RainforthCornishYangWarringtonWood:2018:1}, but they require a large number of forward model evaluations at every design point.
This can get prohibitively expensive during numerical optimization of the EIG, since many evaluations of the EIG are required.
Defining tractable estimators for other commonly-used utility functions is also non-trivial.
For example, Monte Carlo (MC) estimators of the Bayesian A-optimality criterion would require approximating the posterior covariance for \emph{each} data sample used in the MC estimator.

\textbf{Our approach.}\label{page:our-approach}
We develop a black-box approach for optimal experimental design for problems that involve one or more of the following: nonlinear dependence on the parameters and/or the designs, non-differentiable PTO maps, and non-Gaussian priors.
The key to our approach is designing computationally tractable sample-based approximations to the expected utility function.
This is achieved by building a transport map that pushes forward a tractable reference density (\eg, multivariate Gaussian) to the intractable \emph{joint} density for the design, observation and inference random variables.
This transport map can be used to obtain independent and identically distributed (\iid) samples for Monte Carlo approximation to the expected utility for any feasible design and various choices of the utility function.
In addition to facilitating OED, the transport map can also be used to probe the posterior after the experiment has been performed.
In contrast to approaches that utilize the Laplace approximation, our method only requires access to an unnormalized density, and is thus most directly comparable to purely sample-based approaches (\eg, nested Monte Carlo).
Additionally, our method is particularly effective for sequential or adaptive design of experiments, as we discuss in~\cref{section:soed_dirt}.

Transport maps have been used for modeling and exploring posterior distributions in Bayesian inverse problems (see, \eg,~\cite{MoselhyMarzouk:2012:1, ParnoMarzouk:2018:1, BaptistaMarzoukZahm:2020:1, BrennanBigoniZahmSpantiniMarzouk:2020:1}), and have also made an appearance in the OED literature (see, \eg,~\cite{Huan:2015:1, BaptistaCaoChenGhattasLiMarzoukOden:2022:1}).
In~\cite{BaptistaCaoChenGhattasLiMarzoukOden:2022:1}, sample-based transport maps are used to approximate the information contained in various summary statistics of image data, with the ultimate goal of reducing data volume.
In contrast, our focus is choosing experimental conditions that maximize information content of sparse data.
Since the data distribution in our applications depends on the design, we learn the joint density for the data, inference parameter, and design random variables.
This enables approximation of the expected utility function for various designs using a single transport map.
In~\cite{Huan:2015:1}, a transport map to the joint density on the designs, observations and inference parameters is also constructed, but our approach differs in the following key ways:
\begin{enumerate*}[label=\ensuremath{(\roman*)}]
	\item
		we employ a functional tensor train (FTT)-based transport map following~\cite{CuiDolgovZahm:2023:1} and thus take a function approximation viewpoint rather than a sample-based density approximation approach;

	\item
		we target a wide class of Bayesian experimental design problems and optimality criteria, whereas the aforementioned work deals primarily with finding sequential optimal designs (SOEDs) maximizing expected information gain.
\end{enumerate*}

\textbf{Contributions.}
The main contributions of the work presented here are as follows:
\begin{enumerate*}[label=\ensuremath{(\roman*)}]
	\item
		We formulate a computationally tractable transport map approach for approximating expected utility functions appearing in Bayesian OED problems.
		The approach presented is applicable to a wide range of optimality criteria, design types, and prior models, and is thus highly flexible.

	\item
		We extend the base methodology for finding sequentially optimal designs in a greedy fashion.
		To make the sequential OED procedure computationally feasible, we propose recycling previously learned information into suitably chosen preconditioners to speed up subsequent computations.

	\item
		We present a numerical study that illustrates the effectiveness and flexibility of the proposed approach for two types of design problems.
\end{enumerate*}

\textbf{Limitations and outlook.}
Of course, the approach is not without limitations:
\begin{enumerate*}[label=\ensuremath{(\roman*)}]
	\item
		While the approach works well in practice for problems with a \emph{moderate} number of inference parameters, the presented algorithms are typically infeasible for very high- or infinite-dimensional problems.
		This can be circumvented with a-priori dimensionality reduction using, \eg, an extension of the reparametrization techniques presented in~\cite[Section~3]{CuiDolgovZahm:2023:1}.

	\item
		Likewise, the approach works well for finding optimal designs for a small to moderate number of experiments, \ie, in the small data regime, which is the primary focus in this article.

	\item
		It is clear from the design comparisons presented in~\cref{section:NumericalExamples} that the approach seems to work well in practice, however, a theoretical study of the approximation error and rank bounds for the FTT-based surrogates will be crucial to garner a better understanding of the suitability and limitations of our approach.
		First steps to obtain such theoretical guarantees for Gaussian densities can be found in~\cite{RohrbachDolgovGrasedyckScheichl:2022:1}.
\end{enumerate*}

\textbf{Outline.}
In~\cref{section:bground} we outline relevant prerequisite material on the Bayesian approach to inverse problems and Bayesian optimal experimental design (BOED).
In~\cref{section:transport_OED} we use the Knothe-Rosenblatt rearrangement for defining transport-map-based approximations to OED objective functions.
A tractable tensor train construction of the transport maps (building on~\cite{CuiDolgovZahm:2023:1}) is presented in~\cref{section:TT_construction}.
~\Cref{section:soed_dirt} extends our method for guiding data acquisition in a sequential fashion, \ie, for adaptive OED.
In~\cref{section:NumericalExamples}, we illustrate our method with a few numerical examples.

\section{Background}
\label{section:bground}

In \cref{subsection:bground:bayes}, we present some relevant preliminary material on Bayesian inverse problems characterized by non-Gaussian posteriors. We outline the basics of optimal experimental design for such problems as well as the related computational challenges in~\cref{subsection:bground:oed}.

\subsection{Bayesian inverse problems}
\label{subsection:bground:bayes}

We consider the inverse problem of inferring an unknown\breakslash unobservable vector of parameters $\bm \in \mSpace$ from observations of a quantity $\bd \in \dSpace$, related to $\bm = [m_1, m_2, \ldots, m_\Nm]^\transp$ through the model:
\begin{equation}
	\label{eq:invProb_Model}
	\bd
	=
	\cF(\bm) + \boldeta
	,
\end{equation}
where $\cF \colon \mSpace \to \dSpace$ denotes the parameter-to-observable (PTO) map and $\boldeta \in \R^\Nd$ refers to the measurement noise.
The problem-dependent parameter space $\mSpace \subseteq \R^{\Nm}$ and observation space $\dSpace \subseteq \R^{\Nd}$ are assumed to be finite-dimensional.
Additionally, we make the common assumption of an additive Gaussian noise model, though our method could be extended to incorporate other noise models.
In many typical applications, $\cF \coloneqq \cO \circ \cG$ is defined as the composition of a parameter-to-state map $\cG \colon \R^\Nm \to \cU$, where $\cU$ can denote a finite-dimensional or infinite-dimensional vector space, and an observation operator $\cO \colon \cU \to \R^\Nd$.
The observation operator maps the state~$u \in \cU$ to the observable quantity $\bd = [d_1, d_2, \ldots, d_\Nd]^\transp$.
For example, in the simplest case, $\cO$ can denote a (smoothed) point evaluation operator.
Of particular interest in this work are inverse problems where the parameter-to-state operator is expensive to apply, \eg, the parameter-to-state or \emph{forward map} could be defined implicitly through the solution of a PDE or a system of ODEs.
In this case, the parameter vector~$\bm$ typically parametrizes some functional input to the forward operator.

The Bayesian approach is a probabilistic approach to solving inverse problems.
Given a prior probability measure on~$\bm$, denoted by~$\mu_\bm$ with corresponding Lebesgue density~$\pi_\bm$, and mean-zero additive Gaussian noise, $\boldeta \sim \normal{\boldsymbol{0}}{\Gamma_{\boldeta}}$, the solution to the Bayesian inverse problem is a posterior probability law.
In the finite-dimensional parameter inference setting, the posterior law has a corresponding posterior density that can be obtained using Bayes' law,
\begin{equation}
	\label{eq:post_Bayes}
	\pi_{\bm \given \bd}
	=
	\frac{\pi_{\bd \given \bm} \, \pi_\bm}{\pi_\bd}
	.
\end{equation}
The conditional density $\pi_{\bd \given \bm}$, the so-called likelihood, satisfies {$\pi_{\bd \given \bm}(\bd \given \bm) \propto \exp \paren[big](){-\frac{1}{2} \norm{\cF(\bm) - \bd}_{\bGamma^{-1}_{\boldeta}}^2}$} due to the assumption of additive Gaussian measurement noise.
Throughout, for any symmetric positive-definite matrix $\bW \in \R^{n \times n}$ we use $\norm{\,\cdot\,}_\bW$ to denote the $\bW$-weighted norm, \ie, for any $\bx \in \R^n, \norm{\bx}_\bW^2 \coloneqq \bx^\transp \bW \bx$.
The so-called evidence $\pi_\bd$ is typically unknown.

We note that for Bayesian inverse problems where the parameter $\bm$ enters nonlinearly into $\cF$, the posterior $\pi_{\bm \given \bd}$ is typically non-Gaussian and often there is no closed-form expression for the density (even under the assumption of Gaussian prior and additive Gaussian noise).
In such cases, one can only probe the posterior distribution, \eg, by computing various moments or statistics like the maximum a-posteriori estimator (commonly referred to as the MAP point) using sampling techniques such as Markov chain Monte Carlo (MCMC)~\cite{Hastings:1970:1}.
A similar situation arises for linear PTO maps when the prior is non-Gaussian.

\subsection{Optimal experimental design with non-Gaussian posteriors}
\label{subsection:bground:oed}

The quality of the solution to the Bayesian inverse problem depends crucially on the quality and quantity of the measured data.
Guiding data acquisition or choosing experimental conditions for \enquote{optimal} inference of the unknown parameters~$\bm$ requires solving an optimal experimental design problem.
The definition of \enquote{experimental design} is typically problem-specific.
For example, in sensor placement problems, the design could correspond to the spatial coordinates at which the state~$u$ is observed.
Alternatively, the design could enter the PTO map $\cG$ \emph{intrusively} as a boundary\breakslash initial condition, making the state design-dependent.

In the following, we assume that the design can be expressed using the variable $\be \in \eSpace \subset \R^\Ne$ and enters into the model through the parameter-to-observable map ($\cF \colon \eSpace \times \mSpace \rightarrow \dSpace$) as well as the noise model ($\boldeta \sim \normal{\boldsymbol{0}}{\bGamma_{\boldeta}(\be})$).
This includes both intrusive and non-intrusive designs.
The design dependence of the model {and thus of the likelihood} $\pi_{\bd \given \be, \bm}(\bd \given \be, \bm) \propto \exp \paren[big](){-\frac{1}{2} \norm{\cF(\be, \bm)-\bd}_{\Gnoise(\be)^{-1}}^2}$ leads to a design-dependent posterior distribution with density
\begin{equation}
	\pi_{\bm \given \be, \bd}
	=
	\frac{\pi_{\bd \given \be, \bm} \, \pi_\bm}{\pi_{\bd \given \be}}
	,
	\label{eq:posterior_wDesign}
\end{equation}
where we employed the common assumption that prior and design are independent, \ie, $\pi_{\bm \given \be} \equiv \pi_{\bm}$.

Similar to the definition of design, what defines an \enquote{optimal} design is also problem-specific.
For Bayesian inverse problems with non-Gaussian posteriors, optimal designs are often chosen to optimize some expected utility function, \ie,
\begin{equation}
	\be^*
	\in
	\Argmax_{\be \in \eSpace} \Epidge{\psi(\be, \bd)}
	.
	\label{eq:Eutility}
\end{equation}
For any design $\be \in \eSpace$ and corresponding measured data $\bd \in \dSpace$, the utility function $\psi(\be,\bd)$ evaluates the quality of the solution to the resulting Bayesian inverse problem.
This risk-neutral formulation chooses designs that maximize the user-specified utility function on average for all possible realizations of the data.
There are many options for the utility function leading to different optimal designs.
For illustrative purposes we focus on two commonly used criteria in the Bayesian OED literature: A- and D-optimality.

The \emph{A-optimality} criterion seeks designs that minimize the expected value of the average posterior variance in the inference parameters.
This is equivalent to maximizing the expected value of the A-optimal utility function $\psiA$, defined as
\begin{equation}
	\psiA(\be, \bd)
	\coloneqq
	- \trace \paren[big][]{\bC_{\bm \given \be, \bd}(\be, \bd)}
	.
	\label{eq:A-optimality}
\end{equation}
Here $\bC_{\bm \given \be, \bd}(\be, \bd) = \Epimgde[Big]{\paren[big](){\bm - \mpost^{\be, \bd}} \paren[big](){\bm - \mpost^{\be, \bd}}^\transp}$ denotes the posterior covariance matrix with posterior mean $\mpost^{\be, \bd} \coloneqq \Epimgde{\bm}$.

On the other hand, the Bayesian \emph{D-optimal} design maximizes the \emph{expected information gain} (EIG) from prior to posterior (see, \eg,~\cite{Lindley:1956:1}).
Thus the D-optimal utility function $\psiD$ is defined as the Kullback-Leibler (KL) divergence of the posterior from the prior,
\begin{equation}
	\psiD(\be, \bd)
	\coloneqq
	\distKL{\pi_{\bm \given \be, \bd}}{\pi_\bm}
	=
	\Epimgde[Big]{\log \paren[Big](){\frac{\pi_{\bm \given \be, \bd}}{\pi_\bm}}}
	.
	\label{eq:D-optimality}
\end{equation}

We emphasize that for non-Gaussian posteriors, there are typically no closed-form expressions for the utility functions~$\psiA$ and~$\psiD$.
Thus, it is a significant challenge to evaluate these criteria and their expected values, denoted respectively by~$\EpsiA$ and~$\EpsiD$.
To address the latter, we take a sample average approximation (SAA) approach, which leads to the discretized A- and D-optimality criteria:
\begin{align}
	\EpsiA(\be)
	&
	\approx
	- \frac{1}{N} \sum_{i=1}^N \trace \paren[big][]{\bC_{\bm \given \be, \bd^{(i)}}(\be, \bd^{(i)})}
	\eqqcolon
	\PsiA(\be)
	\label{eq:A-objectiveSAA}
	\\
	\EpsiD(\be)
	&
	=
	\Epidmge[Big]{\log \paren[Big](){\frac{\pi_{\bm \given \be, \bd}}{\pi_\bm}}}
	\approx
	\frac{1}{N} \sum_{i=1}^N \log \paren[bigg](){\frac{\pi_{\bm \given \be, \bd^{(i)}}(\bm^{(i)})}{\pi_\bm(\bm^{(i)})}}
	\eqqcolon
	\PsiD(\be)
	,
	\label{eq:D-objectiveSAA}
\end{align}
where $\bd^{(i)} \sim \pi_{\bd \given \be}$ in~\eqref{eq:A-objectiveSAA} and $\paren[big](){\bd^{(i)}, \bm^{(i)}} \sim \pi_{\bd, \bm \given \be}$ in~\eqref{eq:D-objectiveSAA}.
The first equality in~\eqref{eq:D-objectiveSAA} is obtained via another application of Bayes' law.

However, the aforementioned challenge of evaluating the summands remains.
Exact evaluation of each summand is generally infeasible, and some efficient approximation is needed.
These approximations typically require access to the posterior distribution for many realizations of the data and design.
While efficient MCMC methods for obtaining posterior samples and thus exploring intractable distributions have been developed in recent years (see, \eg,~\cite{DodwellKetelsenScheichlTeckentrup:2015:1, CuiLawMarzouk:2016:1}), using these approaches to solve the OED problem~\eqref{eq:A-objectiveSAA} would require generating a different set of samples at each design for each data sample in an iterative optimization algorithm, which can get prohibitively expensive and leads to noisy evaluations of the utility function.

Instead, we employ a measure-transport approach that enables rapid sampling from the joint density~$\pi_{\bd, \bm \given \be}$, the marginal~$\pi_{\bd \given \be}$, and posterior~$\pi_{\bm \given \bd,\be}$ for any fixed design.
The idea, which builds on~\cite{CuiDolgovZahm:2023:1}, is to construct a deterministic coupling between a product-form reference distribution $\nu_{\be, \bd, \bm}$ and the joint distribution for the design, observational and inference random variables $\mu_{\be, \bd, \bm}$.
A particular choice of this coupling, or transport map, called the Knothe-Rosenblatt rearrangement, \enquote{exposes} conditional densities and is thus crucial for our approach.

\section{Optimal experimental design using transport maps}
\label{section:transport_OED}

In this section we present a flexible transport map approach to OED for a wide class of Bayesian inverse problems.
A crucial tool for realizing our approach is the Knothe-Rosenblatt rearrangement, which we define in~\cref{subsection:bground:KRMaps}.
In~\cref{subsection:KR_OED} and~\cref{subsection:OED_u_KR}, we outline how the KR map can be used to approximate the OED objective function~\eqref{eq:Eutility} with particular emphasis on A- and D-optimality.
Error bounds are discussed in~\cref{subsection:errors} with an emphasis on the D-optimal design.

\subsection{Conditional sampling via Knothe-Rosenblatt transport}
\label{subsection:bground:KRMaps}

Given a target random variable $\bx = [x_1, \ldots, x_n]^\transp \in \R^n$ and a reference random variable $\bv = [v_1, \ldots, v_n]^\transp \in \R^n$ with probability laws $\mu_\bx$ and $\nu_\bv$, respectively, the \emph{Knothe-Rosenblatt} (KR) rearrangement~(\cite{Rosenblatt:1952:1, Villani:2009:1}) defines a triangular, order-preserving diffeomorphism $\cT \colon \R^n \to \R^n$ that couples $\bx$ and $\bv$.
The general structure of~$\cT$ is
\begin{equation}
	\cT(\bv)
	=
	\begin{bmatrix}
		x_1
		\\
		x_2
		\\
		\vdots
		\\
		x_n
	\end{bmatrix}
	=
	\begin{bmatrix*}[l]
		\cT_{x_1}(v_1)
		\\
		\cT_{x_2 \given x_1}(\bv_{1:2})
		\\
		\quad
		\vdots
		\\
		\cT_{x_n \given \bx_{1:n-1}}(\bv)
	\end{bmatrix*}
	,
	\label{eq:KR_map_x}
\end{equation}
where the notation $\bx_{1:k} = [x_1,x_2, \ldots, x_k]$ is used to denote the first $k$~components of~$\bx$.
Note that by \emph{order-preserving}, we mean that each component $\cT_{x_k \given \bx_{1:k-1}} \colon \R^k \to \R$ is strictly monotonically increasing in the last variable, $v_k$.
Under this map, the \emph{pushforward} of $\nu_\bv$ is the law of the image $\cT(\bv)$ and is denoted by $\pushforward{\cT} \nu_\bv$.
The \emph{pullback} of $\mu_\bx$ is the law of $\cT^{-1}(\bx)$ denoted by $\pullback{\cT} \mu_\bx$.
In other words, if $\bv \sim \nu_{\bv}$, then $\cT(\bv) \sim \mu_{\bx}$.
Thus, $\cT$ provides a means of transforming samples distributed according to the reference measure $\nu_{\bv}$ to samples distributed according to the target measure $\mu_{\bx}$.

We only consider measures that are absolutely continuous with respect to the Lebesgue measure.
A unique KR rearrangement exists in such cases, and the pushforward and pullback operators define transformations between the target density $\pi_\bx$ (corresponding to $\mu_\bx$) and the reference density $\rho_\bv$ (corresponding to $\nu_\bv$) via the change-of-variables formulae
\begin{align}
	\pi_\bx(\bx)
	&
	=
	\pushforward{\cT} \rho_\bv(\bx)
	=
	\rho_\bv(\cT^{-1}(\bx)) \, \det \paren[big](){D \cT^{-1}(\bx)}
	,
	\\
	\rho_\bv(\bv)
	&
	=
	\pullback{\cT} \pi_\bx(\bv) = \pi_\bx(\cT(\bv)) \, \det \paren[big](){D \cT(\bv)}
	,
	\label{eq:pullback_pushforward}
\end{align}
where $D$ denotes the Jacobian of the respective map. Note that as a consequence of these formulae,
\begin{equation}
	\expect{\pi_{\bx}}{g}
	=
	\expect{\rho_{\bv}}{g\circ \cT}
	\label{eq:COV_exp}
\end{equation}
for any measurable function $g$ on $\R^n$.
Henceforth we assume $\rho_{\bv}$ is a product-form reference density, \ie, it can be written as the product of its marginals, $\rho_{\bv} = \prod_{i=1}^{n} \rho_{v_i}(v_i)$.

Not only does the KR rearrangement offer computational benefits due to its triangular structure, the map also \enquote{exposes} particular conditionals of the density $\pi_{\bx}$.
Let $\bx = (\by,\bz)$ be a partition of the target random variable with $\by \in \R^{n_y}$, $\bz \in \R^{n_z}$ and $n = n_y+n_z$.
Then~\eqref{eq:KR_map_x} can be written as
\begin{equation}
	\cT(\bv)
	\coloneqq
	\cT(\bv_{y},\bv_{z})
	=
	\begin{bmatrix*}[l]
		\cT_{\by}(\bv_{y})
		\\
		\cT_{\bz \given \by}(\bv_{y},\bv_{z})
	\end{bmatrix*}
	,
	\label{eq:KR_conditional}
\end{equation}
where $\bv_{y} \in \R^{n_y}$, $\bv_{z} \in \R^{n_z}$, $\cT_{\by} \colon \R^{n_y} \rightarrow \R^{n_y}$ and $\cT_{\bz \given \by} \colon \R^{n_y} \times \R^{n_z} \rightarrow \R^{n_z}$.
For any fixed $\by^* \in \R^{n_y}$, the map $\cT^{\by^*} \colon \R^{n_z} \rightarrow \R^{n_z}$ (defined as $\bv_{z} \mapsto \cT_{\bz \given \by}(\cT_{\by}^{-1}(\by^*),\bv_{z})$) prescribes a coupling between the marginal reference density $\rho_{\bv_{z}}$ and the conditional target density $\pi_{\bz \given \by = \by^*}$, \ie $\paren[auto](){\cT^{\by^*}}_{\sharp}\rho_{\bv_{z}} = \pi_{\bz \given \by = \by^*}$ \cite[Lemma~1]{MarzoukMoselhyParnoSpantini:2016:2}.

This property makes the KR rearrangement particularly useful for conditional sampling, and we will exploit this feature for our OED approach. However, construction of the true map $\cT$ is in general not feasible and various approximation techniques have been developed.
The most common technique is to define $\cT$ as the minimizer of some statistical divergence between the target density $\pi_{\bx}$ and the pushforward of the reference $\cT_{\sharp}\rho_{\bv}$ over some parametrized class of triangular transport maps.
These techniques use polynomials (\cite{MoselhyMarzouk:2012:1,BaptistaMarzoukZahm:2020:1}) or kernel functions (\cite{LiuWang:2016:1,DetommasoCuiMarzoukSpantiniScheichl:2018:1}) to approximate $\cT$, as well as invertible neural networks and normalizing flows (\cite{KruseDetommasoKoetheScheichl:2019:1,BaptistaHosseiniKovachkiMarzouk:2020:1,PapamakariosNalisnickJimenezRezendeMohamedLakshminarayanan:2021:1}).
In contrast, our approach uses the techniques outlined in~\cite{DolgovAnayaIzquierdoFoxScheichl:2019:1,CuiDolgov:2021:1,CuiDolgovZahm:2023:1} and constructs an explicit transport map to an approximation of the target density, but we delay those details to~\cref{section:TT_construction} and first discuss how the KR rearrangement could be used for approximating the OED objective function.

\subsection{Knothe-Rosenblatt rearrangement for OED}
\label{subsection:KR_OED}

At the core of our approach is the approximation of the joint density  $\piedm$ for the design, observable and inference parameters by the pushforward of a product-form reference density $\rhoedm(\bv) = \rho_\be(v_\be) \, \rho_\bd(v_\bd) \, \rho_\bm(v_\bm)$ with respect to a KR transport map $\cT$.
Using properties of conditional probability distributions and Bayes' law,
\begin{equation}
	\piedm(\be, \bd, \bm)
	=
	\pi_{\bd \given \be, \bm}(\bd \given \be, \bm) \, \pi_\bm(\bm) \, \pi_\be(\be)
	,
	\label{eq:piedm}
\end{equation}
where $\pi_{\bd \given \be, \bm}$ is the design-dependent likelihood, $\pi_\bm$ is the prior, $\pi_\be$ is a user-specified density for the design parameters~$\be$, and $z > 0$ is a normalization constant.
For the remainder of this section, we assume that such a map is available and define $\widehat{\pi}_{\be,\bd,\bm} \coloneqq \pushforward{\cT} \rhoedm$ to be the approximation to the target density $\pi_{\be,\bd,\bm}$.

\begin{remark}[{On the probability density for the designs}]
	\label{remark:pie}
	We note that the marginal density on the designs $\pi_{\be}$ is introduced as a consequence of the decomposition of the joint density $\pi_{\be,\bd,\bm}$ into the product of conditional marginal densities.
	While it is uncommon to encounter an explicit marginal density for the designs in BOED literature, the posterior distribution for~$\bm$ is commonly defined as being conditioned on the design as well as the data, implying that the design is implicitly treated as a random variable.
	Choosing a density~$\pi_{\be}$ can be viewed as enforcing constraints on the design space, hence a natural choice is to weight all the designs in~$\eSpace$ equally by choosing a uniform distribution.
	This is the choice we make for our experiments in~\cref{section:NumericalExamples}.
	However, under our assumptions on the likelihood and prior, the posterior density is independent of the particular choice for~$\pi_{\be}$, thus there is some freedom in choosing~$\pi_{\be}$ and other choices may be conceivable.
	It is important, however, that the support of~$\pi_\be$ covers the space of possible designs~$\eSpace$ and we assume this henceforth.
	Additionally, while the posterior is unchanged for different choices of $\pi_{\be}$, it does of course change the joint density.
\end{remark}

The Knothe-Rosenblatt rearrangement for an arbitrary law on the random variables $\bx = (x_1, \ldots, x_n)$ is unique once a particular arrangement of the variables is prescribed.
While there are advantages to other arrangements of the variables (including ease of computation), we order the design, observable and inference parameter random variables as $\bx = (\be, \bd, \bm)$.
The triangular KR map~$\cT$ is then defined component-wise;  analogously to~\eqref{eq:KR_conditional}, \ie,
\begin{equation}
	\cT(\bv)
	\coloneqq
	\cT(\bv_{e},\bv_{d},\bv_{m})
	=
	\begin{bmatrix*}[l]
		\cT_\be(\bv_e)
		\\
		\cT_{\bd \given \be}(\bv_e, \bv_d)
		\\
		\cT_{\bm \given \be, \bd}(\bv_e, \bv_d, \bv_m)
	\end{bmatrix*}
	=
	\begin{bmatrix}
		\be
		\\
		\bd
		\\
		\bm
	\end{bmatrix}
	=
	\bx
	.
	\label{eq:KR_map_edm}
\end{equation}
The main advantage of using this particular arrangement over alternative orderings is the immediate access to the conditional distributions needed for defining the OED optimality criteria.

Specifically, as described in~\cref{subsection:bground:KRMaps}, for any fixed $\be^* \in \R^{\Ne}$ and $\bd^* \in \R^{\Nd}$, couplings between the approximate evidence density $\widehat{\pi}_{\bd \given \be^*} \approx \pi_{\bd \given \be^*}$ and the marginal reference $\rho_{\bd}$ as well as the approximate posterior density $\widehat{\pi}_{\bm \given \be^*,\bd^*} \approx \pi_{\bm \given \be^*,\bd^*}$ and the marginal reference $\rho_{\bm}$ can be defined through the operators $\cT^{\be^*} \colon \R^{\Nd} \rightarrow \R^{\Nd}$ and $\cT^{\be^*,\bd^*} \colon \R^{\Nm} \rightarrow \R^{\Nm}$ as follows:
\begin{equation}
	\begin{aligned}
		\bv_d
		\sim
		\rho_{\bd}
		&
		\quad
		\Rightarrow
		\quad
		\cT^{\be^*}(\bv_d)
		\coloneqq
		\cT_{\bd\given\be}(\bv_e^*,\bv_d)
		\sim
		\widehat{\pi}_{\bd \given \be^*}
		,
		\\
		\bv_m
		\sim
		\rho_\bm
		&
		\quad
		\Rightarrow
		\quad
		\cT^{\be^*,\bd^*}(\bv_m)
		\coloneqq
		\cT_{\bm \given \be, \bd}(\bv_e^*, \bv_d^*, \bv_m)
		\sim
		\widehat{\pi}_{\bm \given \be^*, \bd^*}
		,
	\end{aligned}
	\label{eq:KR_conditionals}
\end{equation}
where $\bv_e^* \coloneqq \cT^{-1}_\be(\be^*)$ and $\bv_d^* \coloneqq \cT^{-1}_{\bd \given \be}(\be^*, \bd^*)$; in the latter the inverse is taken \wrt the data variable~$\bd$.

\subsection{Approximation of optimality criteria using KR maps}
\label{subsection:OED_u_KR}

Recall that our OED objective is to find a design $\be^* \in \eSpace$ that maximizes an expected utility function~$\Psi$,
\begin{equation*}
	\be^*
	\in
	\Argmax_{\be \in \eSpace} \Psi(\be)
	=
	\Argmax_{\be \in \eSpace} \Epidge{\psi(\be, \bd)}
	.
\end{equation*}
The most evident way to make use of the transport-map-based surrogate for the joint density~$\piedm$ is a direct replacement of all the conditional densities in~$\Psi$ with their corresponding transport-map-based surrogates.
The resulting approximate optimality criterion $\widehat{\Psi}$ can then be approximated using Monte Carlo or Quasi-Monte Carlo quadrature.

This approach could be used for a wide variety of optimality criteria.
Let $\widehat{\psi}$ denote a problem-dependent approximation to the utility function $\psi$ and assume $\bv_{\bd}^{(i)} \sim \rho_{\bd}$.
Using~\eqref{eq:COV_exp}, the procedure for an arbitrary utility function can be summarized with the following sequence of approximations:
\begin{equation*}
	\Psi(\be)
	\approx
	\Epdge[big]{\widehat{\psi}(\be, \bd)}
	=
	\underbrace{\expect[big]{\rho_{\bd}}{\widehat{\psi}(\be, \cT^{\be}(\bv_d))} \vphantom{\sum_{i=1}^N}}_{\eqqcolon \widehat{\Psi}(\be)}
	\approx
	\underbrace{\frac{1}{N}\sum_{i=1}^N \widehat{\psi}(\be, \cT^{\be}(\bv_d^{(i)}))}_{\eqqcolon \widehat{\Psi}^N(\be)}
	.
\end{equation*}
However, in the results presented in~\cref{section:NumericalExamples}, we focus on the commonly-used A- and D-optimal design objectives.
Hence, we specify the exact procedure used for approximating those two objectives.

\subsection*{A-Optimal design}

We propose the following approximation to the A-optimal objective for any $\be \in \eSpace$:
\begin{equation}
	\begin{aligned}
		\EpsiA(\be)
		=
		- \Epidge[big]{\trace \paren[big](){\bC _{\bm \given \be, \bd}(\be, \bd)}}
		&
		\approx
		- \Epdge[big]{\trace \paren[big](){\widehat{\bC}_{\bm \given \be, \bd}(\be, \bd)}} \eqqcolon \EhatPsiA(\be)
		\\
		&
		\approx
		- \frac{1}{N} \sum_{i=1}^N \trace \paren[big](){\widehat{\bC}_{\bm \given \be, \bd^{(i)}}(\be, \bd^{(i)})}
		\eqqcolon
		\PsiA(\be)
		.
	\end{aligned}
\end{equation}
Here, $\{\bd^{(i)}\}_{i=1}^{N}$ are data samples from the approximate evidence, $\bd^{(i)} \sim \widehat{\pi}_{\bd \given \be}$, and the covariance matrix for each data sample $\bC _{\bm \given \be, \bd^{(i)}}$ is approximated via a sample average as
\begin{equation*}
	\bC _{\bm \given \be, \bd^{(i)}}(\be, \bd^{(i)})
	\approx
	\widehat{\bC}_{\bm \given \be, \bd^{(i)}}(\be, \bd^{(i)})
	\coloneqq
	\frac{1}{M-1} \sum_{k=1}^M\paren[big](){\bm^{(k,i)} - \overline{\bm}^{(i)}}\paren[big](){\bm^{(k,i)} - \overline{\bm}^{(i)}}^\transp
\end{equation*}
with $\bm^{(k,i)} \sim \widehat{\pi}_{\bm \given \be, \bd^{(i)}}$ and $\overline{\bm}^{(i)} \coloneqq \frac{1}{M} \sum_{k=1}^M \bm^{(k,i)}$.
The samples $\bd^{(i)}$ and $\bm^{(k,i)}$ are obtained using the transport map $\cT$ as described in~\eqref{eq:KR_conditionals}.
The full procedure for evaluating $\PsiA(\be)$ for any design $\be \in \eSpace$ is summarized in~\cref{algorithm:AOpt}.
\begin{algorithm}[h!]
	\caption{Evaluate $\PsiA(\be) \approx \EpsiA(\be)$ using transport map to joint density.}
	\label{algorithm:AOpt}
	\begin{algorithmic}[1]
		\Procedure{AOpt}{$\rhoedm, \cT, \be, N, M$}
		\Comment{$\pushforward{\cT} \rhoedm \approx \piedm$}
		\State Sample $\bv_d^{(i)} \sim \rho_\bd$ for $i=1, \ldots, N$, $\bv_m^{(k)} \sim \rho_\bm$ for $k = 1, \ldots, M$
		\State $\bd^{(i)} \gets \cT^{\be}(\bv_d^{(i)})$ for $i=1, \ldots, N$
		\Comment{\cref{eq:KR_conditionals}}
		\For{i=1, \ldots, N}
		\State $\bm^{(k,i)} \gets \cT^{\be, \bd^{(i)}}(\bv_{\bm}^{(k)})$ for $k = 1, \ldots, M$
		\Comment{\cref{eq:KR_conditionals}}
		\State $\overline{\bm}^{(i)} \gets \frac{1}{M}\sum_{k=1}^M \bm^{(k,i)}$
		\State $\widehat{\bC}_{\bm \given \be, \bd^{(i)}}(\be, \bd^{(i)})
		\gets
		\frac{1}{M-1} \sum_{k=1}^M \paren[big](){\bm^{(k,i)} - \overline{\bm}^{(i)}} \paren[big](){\bm^{(k,i)} - \overline{\bm}^{(i)}}^\transp$
		\EndFor
		\State $\PsiA \gets \frac{1}{N} \sum_{i=1}^N \trace \paren[big][]{\widehat{\bC}_{\bm \given \be, \bd^{(i)}}(\be, \bd^{(i)})}$
		\State \Return $\PsiA(\be)$
		\EndProcedure
	\end{algorithmic}
\end{algorithm}

\subsection*{D-Optimal design}

Similarly, we propose the following approximation to the expected information gain (EIG):
\begin{equation}
	\begin{aligned}
		\EpsiD(\be)
		=
		\Epidmge[Big]{\log \paren[Big](){\frac{\pi_{\bm \given \be, \bd}}{\pi_\bm}}}
		&
		\approx
		\Epdmge[Big]{\log \paren[Big](){\frac{\widehat{\pi}_{\bm \given \be, \bd}}{\pi_\bm}}} \eqqcolon \EhatPsiD(\be)
		\\
		&
		\approx
		\frac{1}{N} \sum_{i=1}^N \log \paren[bigg](){\frac{\widehat{\pi}_{\bm \given \be, \bd} \paren[big](){\bm^{(i)} \given \be, \bd^{(i)}}}{\pi_\bm(\bm^{(i)})}}
		\eqqcolon
		\PsiD(\be)
		.
	\end{aligned}
	\label{eq:DOpt_TM}
\end{equation}
Here, joint samples $\paren[big](){\bd^{(i)}, \bm^{(i)}} \sim \widehat{\pi}_{\bd, \bm \given \be}$ are obtained by transforming $\paren[big](){\bv_{d},\bv_{m}} \sim \rho_{\bd,\bm}$ using the transport map $\cT$ and~\eqref{eq:KR_conditionals}.
~\Cref{algorithm:EIG} summarizes the procedure for evaluating $\PsiD(\be)$ for any $\be \in \eSpace$.

\begin{algorithm}[ht]
	\caption{Evaluate $\PsiD(\be) \approx \EpsiD(\be)$ using transport map to joint density.}
	\label{algorithm:EIG}
	\begin{algorithmic}[1]
		\Procedure{DOpt}{$\rhoedm, \cT, \pi_\bm, \be,N$}
		\Comment{$\pushforward{\cT} \rhoedm \approx \piedm$}
		\State Sample $\paren[big](){\bv_d^{(i)}, \bv_m^{(i)}} \sim \rho_{\bd, \bm}$ for $i=1, \ldots, N$
		\State $\paren[big](){\bd^{(i)}, \bm^{(i)}} \gets \paren[big](){\cT^{\be}(\bv_{d}^{(i)}),\cT^{\be,\bd^{(i)}}(\bv_{m}^{(i)})}$ for $i=1, \ldots, N$
		\Comment{\cref{eq:KR_conditionals}}
		\For{i=1, \ldots, N}
		\State $\widehat{\pi}_{\bm \given \be, \bd^{(i)}}(\bm^{(i)}) = \pushforward{(\cT^{\be,\bd^{(i)}})} \rho_\bm(\bm^i)$
		\Comment{$\cT^{\be,\bd^{(i)}}$ defined in~\eqref{eq:KR_conditionals}}
		\State $\psiD(\be, \bd^{(i)}, \bm^{(i)}) \gets \log \paren[Big](){\frac{\widehat{\pi}_{\bm \given \be, \bd^{(i)}}(\bm^{(i)})}{\pi_\bm(\bm^{(i)})}}$
		\EndFor
		\State $\PsiD(\be) \gets \frac{1}{N} \sum_{i=1}^N \psiD(\be, \bd^{(i)}, \bm^{(i)})$
		\State \Return $\PsiD(\be)$
		\EndProcedure
	\end{algorithmic}
\end{algorithm}

\bigskip
With these Monte Carlo approximations to~$\Psi_X$ (for $X = A,D$), the OED goal reduces to finding a design vector~$\be^*$ that maximizes the tractable approximate expected utility function
\begin{equation}
	\be^*
	\in
	\Argmax_{\be \in \eSpace} \PsiX(\be)
	.
\end{equation}
A detailed discussion of viable algorithms for optimizing $\PsiX$ is outside the scope of this work.
For the numerical results in~\cref{section:NumericalExamples}, we use gradient-free optimization that do not require of~$\PsiX$.
Of course, gradient-based optimizers can also be used, provided~$\PsiX(\be)$ is differentiable with respect to~$\be$.
This choice is left to be a black box to be specified by the user.

\subsection{Error estimates for the transport-based approximations of expected information gain}\label{subsection:errors}

In the following, we derive probabilistic bounds for the error due to approximating the D-optimality criterion~$\EpsiD$ by the corresponding transport-map-based surrogate~$\EhatPsiD$ defined in~\eqref{eq:DOpt_TM}.
That is, for any $\zeta > 0$, we provide a lower bound for
\begin{equation}
	\prob[auto]{\pi_\be}{\abs[big]{\EpsiD(\be) - \EhatPsiD(\be)} \le \zeta}
	,
	\label{eq:TM_approx_err}
\end{equation}
where $\abs[big]{\EpsiD(\be) - \EhatPsiD(\be)} \coloneqq \abs[big]{\Epidmge{h} - \Epdmge{\widehat{h}}}$ with
\begin{equation}
	\label{eq:h-and-hhat:D-criterion}
	h(\be, \bd, \bm)
	\coloneqq
	\log \paren[auto](){\frac{\pi_{\bm \given \be, \bd}}{\pi_\bm}}
	\quad
	\text{and}
	\quad
	\widehat{h}(\be, \bd, \bm)
	\coloneqq
	\log \paren[auto](){\frac{\widehat{\pi}_{\bm \given \be, \bd}}{\pi_\bm}}
	.
\end{equation}
We focus on the D-optimality criterion here for illustrative purposes.
Bounds for the A-optimality criterion can be derived similarly.

The error estimates we derive depend on the Hellinger distance between the target and surrogate densities $\pi_{\be,\bd,\bm}$ and $\widehat{\pi}_{\be,\bd,\bm}$:
\begin{equation}
	\distH{\piedm}{\pihatedm}
	\coloneqq
	\paren[Big](){\frac{1}{2} \int \paren[auto](){\sqrt{\widehat{\pi}_{\be, \bd, \bm}} - \sqrt{\pi_{\be, \bd, \bm}}}^2 \, \d \bd \d \bm}^{\frac{1}{2}}
	.
\end{equation}
As shown in~\cref{section:TT_construction}, this choice is particularly useful for tensor-train-based approximations to the transport maps, since the resulting surrogates naturally satisfy $\distH{\piedm}{\pihatedm} \le \varepsilon$, for some $\varepsilon \ge 0$.
However, alternative constructions using other statistical distances\breakslash divergences, \eg, $\distKL{\pi_{\be,\bd,\bm}}{\widehat{\pi}_{\be,\bd,\bm}}$ may lead to potentially sharper error estimates using other notions of distance or divergence.

To derive the lower bound, we make use of the following lemmas, the first two of which concerns arbitrary PDFs of a random variable $\bx$ taking values in $\xSpace \subseteq \R^{n_x}$ for some $n_x \in \N$.
To ease readability, we let
\begin{equation*}
	L^2_{\pi}(\xSpace)
	\coloneqq
	\setDef{h \colon \xSpace \rightarrow \R}{\text{$h$ is measurable and $\norm{h}_{L^2_\pi} < \infty$}}
\end{equation*}
where
\begin{equation*}
	\norm{h}_{L^2_\pi}
	\coloneqq
	\sqrt{\expect{\pi}{h^2}}
	.
\end{equation*}
\begin{lemma}[\protect{\cite[Proposition~6]{CuiDolgov:2021:1}}]
	Let $\pi$ and $\widehat{\pi}$ be two PDFs.
	For any function $h \in L^2_{\pi}(\xSpace) \cap L^2_{\widehat{\pi}}(\xSpace)$,
	\begin{equation}
		\abs[big]{\expect{\pi}{h} - \expect{\widehat{\pi}}{h}}
		\le
		\sqrt{2} \, \distH{\pi}{\widehat{\pi}} \paren[Big](){\norm{h}_{L^2_\pi} + \norm{h}_{L^2_{\widehat{\pi}}}}
		.
	\end{equation}
	\label{lemma:CuiDolgov1}
\end{lemma}

\begin{lemma}
	Let the ratio of two PDFs $\pi$ and $\widehat{\pi}$ be bounded from above almost surely, \ie,
	\begin{equation}
		\sup_{\bx \in \xSpace \subseteq \R^{n_x}} \frac{\pi(\bx)}{\widehat{\pi}(\bx)}
		\le
		c
		<
		\infty
		.
		\label{eq:bdd_ratio}
	\end{equation}
	Then the KL divergence of $\pi$ from $\widehat{\pi}$ can be bounded as follows:
	\begin{equation}
		\distKL{\pi}{\widehat{\pi}}
		\le
		\sqrt{2} \, \distH{\pi}{\widehat{\pi}} \paren[auto](){\norm[auto]{\frac{\pi}{\widehat{\pi}}}_{L^2_\pi} + \norm[auto]{\frac{\pi}{\widehat{\pi}}}_{L^2_{\widehat{\pi}}}}
		.
	\end{equation}
	\label{lemma:chiBound}
\end{lemma}
\begin{proof}
	Under our assumptions, using~\cite[Theorem~5]{GibbsSu:2002:1}, $\distKL{\pi}{\widehat{\pi}} \le \distChi{\pi}{\widehat{\pi}}$ where $\distChi{\pi}{\widehat{\pi}}$ denotes the $\chi^2$-divergence of $\pi$ from $\widehat{\pi}$. The statement then follows from~\cite[Corollary~2]{CuiDolgov:2021:1}.
\end{proof}

Now, consider two probability densities $\pi_{\bx,\by}=\pi_{\bx \given \by}\pi_{\by}$ and $\widehat{\pi}_{\bx,\by}=\widehat{\pi}_{\bx \given \by}\widehat{\pi}_{\by}$  defined on the product space $\xSpace \times \ySpace \subseteq \R^{n_x}\times\R^{n_y}$ (with $n_x,n_y \in \N$).

\begin{lemma}[\protect{\cite[Lemma~4]{CuiDolgovZahm:2023:1}}]
	The expected value of the Hellinger distance between $\pi_{\bx \given \by}$ and $\widehat{\pi}_{\bx \given \by}$ satisfies:
	\begin{equation}
		\expect[Big]{\pi_\by}{\distH{\pi_{\bx \given \by}}{\widehat{\pi}_{\bx \given \by}}}
		\le
		2 \, \distH{\pi_{\bx, \by}}{\widehat{\pi}_{\bx, \by}}
		.
		\label{eq:conditionalHell}
	\end{equation}
	Furthermore,
	\begin{equation}
		\expect[Big]{\pi_\by}{\distH{\pi_{\bx \given \by}}{\widehat{\pi}_{\bx \given \by}}^2}
		\le
		4 \, \distH{\pi_{\bx, \by}}{\widehat{\pi}_{\bx, \by}}^2
		.
		\label{eq:conditionalHell2}
	\end{equation}
	\label{lemma:conditionalHell}
\end{lemma}

\begin{proposition}
	Assume the PDFs $\pi_{\bx,\by}$ and $\widehat{\pi}_{\bx,\by}$ satisfy $\distH{\pi_{\bx,\by}}{\widehat{\pi}_{\bx,\by}} \le \varepsilon$ for some $\varepsilon \ge 0$.
	Furthermore, assume that the ratio of the conditional densities $\pi_{\bx \given \by}$ and $\widehat{\pi}_{\bx \given \by}$ is bounded almost surely by $c_1 \in L^2_{\pi_{\by}}(\ySpace)$, \ie, for all $\by \in \ySpace$,
	\begin{equation}
		\sup_{\bx \in \xSpace} \frac{\pi_{\bx \given \by}(\bx)}{\widehat{\pi}_{\bx \given \by}(\bx)}
		\le c_1(\by)
		.
		\label{eq:ratio_pi_p}
	\end{equation}
	Then we have the following bound on the expected error:
	\begin{equation}
		\Epixy[Big]{\log \paren[Big](){\frac{\pi_{\bx \given \by}}{\widehat{\pi}_{\bx \given \by}}}}
		\le
		4 \, \sqrt{2} \, \varepsilon \, \norm[auto]{c_1}_{L^2_{\pi_{\by}}}
		.
		\label{eq:bound_Eintegrand}
	\end{equation}
	\begin{proof}
		Note that
		\begin{align*}
			\Epixy[Big]{\log \paren[Big](){\frac{\pi_{\bx \given \by}}{\widehat{\pi}_{\bx \given \by}}}}
			&
			=
			\Epiy[auto]{\Epixgy[Big]{\log \paren[Big](){\frac{\pi_{\bx \given \by}}{\widehat{\pi}_{\bx \given \by}}}}}
			\\
			&
			=
			\Epiy[big]{\distKL{\pi_{\bx \given \by}}{\widehat{\pi}_{\bx \given \by}}}
			\le
			2 \, \sqrt{2} \, \Epiy[big]{c_1(\by) \, \distH{\pi_{\bx \given \by}}{\widehat{\pi}_{\bx \given \by}}}
			,
		\end{align*}
		where the inequality follows from~\cref{lemma:chiBound} and assumption~\eqref{eq:ratio_pi_p}.
		Then, using Hölder's inequality and~\cref{lemma:conditionalHell}, we have
		\begin{equation*}
			\Epiy[big]{c_1(\by) \, \distH{\pi_{\bx \given \by}}{\widehat{\pi}_{\bx \given \by}}}
			\le
			2 \, \distH{\pi_{\bx,\by}}{\widehat{\pi}_{\bx,\by}} \, \norm[auto]{c_1}_{L^2_{\pi_{\by}}}
			,
		\end{equation*}
		and the result~\eqref{eq:bound_Eintegrand} thus follows.
	\end{proof}
	\label{proposition:bound_Eintegrand}
\end{proposition}

\begin{proposition}
	Let $\pi_{\bx,\by}$ and $\widehat{\pi}_{\bx,\by}$ be two PDFs satisfying $\distH{\pi_{\bx,\by}}{\widehat{\pi}_{\bx,\by}} \le \varepsilon$ for some $\varepsilon \ge 0$.
	Assume that the function $\widehat{h}(\by,\cdot)$ is bounded almost surely from above by $c_2(\by)$ with $c_2 \in L^2_{\pi_{\by}}(\ySpace)$, \ie,
	\begin{equation}
		\sup_{\bx \in \xSpace} \, \abs[big]{\widehat{h}(\by,\bx)}
		\le
		c_2(\by)
		, \quad {\text{for all } \by \in \ySpace}.
		\label{eq:assumption_hBoundedMoments}
	\end{equation}
	Then
	\begin{equation}
		\Epiy[auto]{\abs[big]{\Epixgy{\widehat{h}} - \Epxgy{\widehat{h}}}}
		\le
		4 \, \sqrt{2} \, \varepsilon \, \norm[auto]{c_2}_{L^2_{\pi_{\by}}}
		.
	\end{equation}
	\label{proposition:bound_Epdf}
\end{proposition}
\begin{proof}
	Using~\cref{lemma:CuiDolgov1}, we have that
	\begin{equation*}
		\Epiy[auto]{\abs[big]{\Epixgy{\widehat{h}} - \Epxgy{\widehat{h}}}}
		\le
		\Epiy[auto]{\sqrt{2} \, \distH{\pi_{\bx \given \by}}{\widehat{\pi}_{\bx \given \by}} \paren[Big](){\norm{\hat{h}}_{L^2_{\pi_{\bx \given \by}}} + \norm{\hat{h}}_{L^2_{\widehat{\pi}_{\bx \given \by}}}}}
		.
	\end{equation*}
	The result then follows from~\eqref{eq:assumption_hBoundedMoments},~\cref{lemma:conditionalHell}, and Hölder's inequality.
\end{proof}

Now we are ready to prove the main result of this section, which is a probabilistic bound for the error induced by the use of a transport-map-based approximation to the D-optimality criterion.
\begin{theorem}
	Let $\piedm$ be the joint density for the design ($\be$), observable ($\bd$), and inference parameters ($\bm$), as defined in~\cref{subsection:KR_OED}. Let $\pihatedm$ be some approximation to $\pi_{\be,\bd,\bm}$ with $\distH{\piedm}{\pihatedm} \le \varepsilon$ for some $\varepsilon \ge 0$.
	Furthermore, for all $\be \in \eSpace$ and $\bd \in \dSpace$, assume
	\begin{alignat}{2}
		\sup_{\bm \in \mSpace} \frac{\pi_{\bm \given \be,\bd}(\bm)}{\widehat{\pi}_{\bm \given \be, \bd}(\bm)}
		&
		\le
		c_1(\be,\bd)
		&
		&
		\quad
		\text{for all }
		(\be,\bd) \in \eSpace \times \dSpace
		\text{ and}
		\label{eq:assumption_ratio}
		\\
		\sup_{(\bd, \bm) \in \dSpace \times \mSpace} \, \abs[Big]{\log \paren[Big](){\frac{\widehat{\pi}_{\bm \given \be, \bd}(\bm)}{\pi_\bm(\bm)}}}
		&
		\le
		c_2(\be)
		&
		&
		\quad
		\text{for all }
		\be \in \eSpace
		\label{eq:assumption_log}
	\end{alignat}
	for some $c_1 \in L^2_{\pi_{\be,\bd}}(\eSpace\times\dSpace)$ and $c_2 \in L^2_{\pi_{\be}}(\eSpace)$ (with the spaces $\eSpace$ and $\dSpace$ as defined as in~\cref{section:bground}).
	Then,
	\begin{equation}
		\Epie[auto]{\abs[big]{\EpsiD(\be)-\EhatPsiD(\be)}}
		\le
		4 \, \sqrt{2} \, \varepsilon \, \paren[auto](){\norm{c_1}_{L^2_{\pi_{\be,\bd}}} + \norm{\hat{c_2}}_{L^2_{\pi_{\be}}}}
		.
		\label{eq:Dopt_E_bound}
	\end{equation}
	Additionally, for any $\zeta > 0$,
	\begin{equation}
		\prob[auto]{\pi_\be}{\abs[big]{\EpsiD(\be) - \EhatPsiD(\be)} \le \zeta}
		\ge
		1 - \frac{4 \, \sqrt{2} \, \varepsilon}{\zeta} \paren[auto](){\norm{c_1}_{L^2_{\pi_{\be,\bd}}} + \norm{\hat{c_2}}_{L^2_{\pi_{\be}}}}
		.
		\label{eq:Dopt_prob_bound}
	\end{equation}
	\begin{proof}
		Let $h(\be, \bd, \bm) \coloneqq \log \paren[big](){\frac{\pi_{\bm \given \be, \bd}}{\pi_\bm}}$ and $\widehat{h}(\be, \bd, \bm) \coloneqq \log \paren[big](){\frac{\widehat{\pi}_{\bm \given \be, \bd}}{\pi_\bm}}$.
		By the triangle inequality, we have:
		\begin{align*}
			\Epie[auto]{\abs[big]{\EpsiD(\be)-\EhatPsiD(\be)}}
			&
			=
			\Epie[auto]{\abs[big]{\Epidmge{h} - \Epdmge{\widehat{h}}}}
			\\
			&
			\le
			\Epie[auto]{\abs[big]{\Epidmge{h - \widehat{h}}}}
			+
			\Epie[auto]{\abs[big]{\Epidmge{\widehat{h}} - \Epdmge{\widehat{h}}}}
			\\
			&
			=
			\Epiedm[Big]{\log \paren[Big](){\frac{\pi_{\bm \given \be, \bd}}{\widehat{\pi}_{\bm \given \be, \bd}}}}
			+
			\Epie[auto]{\abs[big]{\Epidmge{\widehat{h}} - \Epdmge{\widehat{h}}}}
			.
		\end{align*}
		The first inequality~\eqref{eq:Dopt_E_bound} thus follows from~\cref{proposition:bound_Eintegrand} (with $\bx = \bm$ and $\by = (\be,\bd)$) and~\cref{proposition:bound_Epdf} (with $\bx = (\bd,\bm)$, $\by = \be$, and $\widehat{h}(\by,\bx) \coloneqq \widehat{h}(\be,\bd,\bm) = \log\paren[auto](){\frac{\widehat{\pi}_{\bm\given\be,\bd}}{\pi_{\bm}}}$).
		The second statement~\eqref{eq:Dopt_prob_bound} can be deduced by applying Markov's inequality.
	\end{proof}
	\label{theorem:ErrorBound}
\end{theorem}

\begin{remark}[Some comments on the bounds and assumptions of~\cref{theorem:ErrorBound}.]
	Note that for better readability, we make use of the constants $c_1(\be, \bd)$ and $c_2(\be)$ to bound the expectations of $\frac{\pi_{\bm \given \be, \bd}}{\widehat{\pi}_{\bm \given \be, \bd}}$ and $\log \paren[auto](){\frac{\widehat{\pi}_{\bm \given \be, \bd}}{\pi_\bm}}$ respectively.
	Tighter upper bounds can be obtained by foregoing this simplification.
	Additionally, note that there is further error incurred due to the use of Monte Carlo approximation in the D-optimality criterion~\eqref{eq:DOpt_TM}, \ie,
	\begin{equation*}
		\Epie[auto]{\abs[big]{\EpsiD(\be)-\PsiD(\be)}}
		\le
		\Epie[auto]{\abs[big]{\EpsiD(\be)-\EhatPsiD(\be)}}
		+
		\Epie[auto]{\abs[big]{\EhatPsiD(\be)-\PsiD(\be)}}
		.
	\end{equation*}
	However, $\Epie[auto]{\abs[big]{\EhatPsiD(\be)-\PsiD(\be)}}$ can be controlled by the sample size~$N$.
\end{remark}

\section{Construction of the Knothe-Rosenblatt rearrangement using tensor trains}\label{section:TT_construction}

In this section, we discuss an efficient numerical realization of the KR rearrangement~\eqref{eq:KR_map_edm}.
Building on~\cite{DolgovAnayaIzquierdoFoxScheichl:2019:1,CuiDolgov:2021:1,CuiDolgovZahm:2023:1}, we construct the transport map via a functional tensor train (FTT) approximation to the square-root of the joint target density.
We outline pertinent background material on FTTs in~\cref{subsection:FTTs} and detail their utility in implementing the KR rearrangement in~\cref{subsection:iRT_TT}.
Finally we present our FTT-based construction of the KR map approximating the joint density $\piedm$ and outline the corresponding OED algorithm in~\cref{subsection:KR_OED_Difficulties}.

\subsection{Functional tensor trains}\label{subsection:FTTs}

The primary computational tool used in our approach is the functional tensor train decomposition~\cite{BigoniEngsigKarupMarzouk:2016:1,GorodetskyKaramanMarzouk:2019:1}.
Given an arbitrary multivariate function $g \colon \xSpace \rightarrow \R$ with $\xSpace = \xSpace_1 \times \xSpace_2 \times \ldots \xSpace_{\Ntot}$, the FTT approximates $g(\bx)$ as the product of $\Ntot$ univariate, matrix-valued functions $\bG_k(x_k)$, \ie,
\begin{equation}
	g(\bx) \approx \widehat{g}(\bx) \coloneqq \bG_1(x_i) \cdots \bG_k(x_k) \cdots \bG_{\Ntot}(x_{\Ntot}).
	\label{eq:FTT}
\end{equation}
The functions $\bG_k \colon \xSpace_k \rightarrow \R^{r_{k-1} \times r_k}$ (with $r_{0} = r_{\Ntot} = 1$) are commonly referred to as \emph{TT cores} and the dimensions $r_k$ are the corresponding \emph{TT ranks}.
Denoting by $\bA_k \in \R^{r_{k-1} \times \Nbasis_k \times r_k}$ a third-order coefficient tensor, the components of each $\bG_k$ can be represented as a linear combination of $\Nbasis_k$ basis functions $\{\basis_k^{(\ell)}\}_{\ell=1}^{\Nbasis_k}$, \ie,
\begin{equation}
	\paren[auto][]{\bG_k(x_k)}_{i,j} = \sum_{\ell=1}^{\Nbasis_k} \basis_k^{(\ell)}(x_k)\bA_k[i,\ell,j]
	,
	\quad
	\text{for }
	i = 1, \ldots, r_{k-1}
	\text{ and }
	j = 1, \ldots, r_k
	.
\end{equation}

The FTT decomposition can be built using efficient alternating-direction TT-cross approximation methods; for details see~\cite{OseledetsTyrtyshnikov:2010:1,DolgovSavostyanov:2014:1}.
We use a functional extension (described in~\cite[Appendix~A]{CuiDolgov:2021:1} and~\cite[Appendix~B]{CuiDolgovZahm:2023:1}) of the rank-adaptive alternating minimal energy method in~\cite{DolgovSavostyanov:2014:1}.
In general, the resulting decomposition only provides an approximation to the multivariate target function.
However, given an error tolerance $\widetilde{\varepsilon} > 0$, this procedure adaptively chooses the TT ranks to minimize the $L^2$ residual error, resulting in an FTT approximation satisfying
\begin{equation}
	\label{eq:FTT-approximation}
	\norm[auto]{g - \widehat{g}}_{L^2}
	\coloneqq
	\paren[auto](){ \int_{\xSpace} \abs{g(\bx)-\widehat{g}(\bx)}^2 \d \bx}^{\frac{1}{2}}
	\le
	\widetilde{\varepsilon}
	.
\end{equation}

Introducing upper bounds on the ranks $r = \max_{k} r_k$ and on the number of basis functions $\Nbasis = \max_{k} \Nbasis_k$, the procedure constructs $\widehat{g}$ with $\cO\paren[auto](){\Ntot \, \Nbasis \, r^2}$ evaluations of the function $g$ and $\cO\paren[auto](){\Ntot \, \Nbasis \, r^3}$ floating point operations (FLOPs).
Thus, for problems where the function $g$ is expensive to evaluate, the efficiency of the algorithm is highly dependent on the maximum ranks required to satisfy the $\widetilde{\varepsilon}$~error tolerance.
The maximum rank typically depends on various properties of the function (smoothness, level of nonlinearity, dimension of $\xSpace$, \etc), hence bounding the maximal rank for general functions is challenging.
However, bounds have been established for some specific types of functions in~\cite{RohrbachDolgovGrasedyckScheichl:2022:1,GriebelHarbrecht:2021:1}.

\subsection{Construction of KR rearrangements via tensor train surrogates}\label{subsection:iRT_TT}

Following~\cite{DolgovAnayaIzquierdoFoxScheichl:2019:1,CuiDolgov:2021:1,CuiDolgovZahm:2023:1,ZhaoCui:2023:1}, we first outline the procedure for constructing the KR rearrangement in the general setting.
Details of its application to design problems are given in the subsequent section.
Specifically, here we consider the random variable $\bx \in \xSpace = \xSpace_1 \times \ldots \times \xSpace_{\Ntot}$ with corresponding density
\begin{equation}
	\pi_{\bx}(\bx) = \frac{p_{\bx}(\bx)}{Z},
	\label{eq:normalizedDensity}
\end{equation}
where $Z > 0$ is the normalizing constant.
Throughout this section, we assume access to the $\emph{unnormalized}$ density $p_{\bx}$ and treat $Z$ as unknown.

Let $\unifDensity_{\bx}$ be a uniform reference density defined on the unit hypercube $[0,1]^{\Ntot}$.
To construct a KR map, coupling $\unifDensity_{\bx}$ and ${\pi}_{\bx}$, we make use of the component notation $\bx_{1:k} = [x_1, \ldots, x_k]$, and let
\begin{align*}
	F_{x_1}(x_1)
	&
	\coloneqq
	\int_{-\infty}^{x_1} {\pi}_{x_1}(x_1') \, \d x_1',
	\\
	F_{x_{k} \given \bx_{1:k-1}}(\bx_{1:k})
	&
	\coloneqq
	\int_{-\infty}^{x_k} {\pi}_{x_k \given \bx_{1:k-1}}(x_k' \given \bx_{1:k-1}) \, \d x_k'
	=
	\int_{-\infty}^{x_k} \frac{{\pi}_{\bx_{1:k}}(\bx_{1:k-1},x_k')}{{\pi}_{\bx_{1:k-1}}(\bx_{1:k-1})} \, \d x_k' \quad \text{for } k = 2, \ldots, \Ntot
	,
\end{align*}
denote cumulative distribution functions (CDFs), where the marginal densities ${\pi}_{\bx_{1:k}}$ (for $k = 1, \ldots, \Ntot-1$) are given by
\begin{equation}
	\pi_{\bx_{1:k}}(\bx_{1:k})
	\coloneqq
	\int_{\xSpace_{>k}} {\pi}_{\bx}(\bx) \, \d \bx_{k+1:\Ntot}
	,
	\quad
	\xSpace_{>k} \coloneqq \xSpace_{k+1}\times\ldots\times\xSpace_{\Ntot}
	.
	\label{eq:marginals}
\end{equation}
Note that the map $\cS \colon \xSpace \to [0,1]^{\Ntot}$, with
\begin{equation}
	\cS(\bx)
	=
	\begin{bmatrix*}[l]
		F_{x_1}(x_1)
		\\
		F_{x_2 \given x_1}(\bx_{1:2})
		\\
		\quad
		\quad
		\vdots
		\\
		F_{x_\Ntot \given \bx_{1:\Ntot-1}}(\bx)
	\end{bmatrix*}
	\eqqcolon
	\bu
	,
\end{equation}
defines a triangular order-preserving coupling (\ie, a KR rearrangement) between the uniform reference and the target, such that $\pushforward{\cS}{\pi}_{\bx} = \unifDensity_{\bx}$.
Through inversion of the map $\cS$, we can define the map $\cT \equiv \cS^{-1} \colon [0,1]^{\Ntot} \rightarrow \xSpace$ that pushes forward the uniform reference to the target, $\pushforward{\cT} \unifDensity_{\bx} = {\pi}_{\bx}$.
Note that for any $\bu \in [0,1]^{\Ntot}$, the inverse map
\begin{equation}
	\cT(\bu)
	\coloneqq
	[F_{x_1}^{-1}(u_1), F_{x_2\given x_1}^{-1}(\bu_{1:2}), \ldots, F^{-1}_{x_{\Ntot}\given \bx_{1:\Ntot-1}}(\bu)]^\transp
	\label{eq:iRT}
\end{equation}
can be evaluated via a sequence of $\Ntot$ inversions of scalar-valued functions.
Specifically, $F^{-1}_{x_k \given \bx_{1:k-1}}(\bu_{1:k})$ is evaluated by inverting $u_k = F_{x_k\given \bx_{1:k-1}}(\bx_{1:k-1},x_k)$ for $x_k$, where $\bx_{1:k-1}$ is determined from the previous $k-1$ inversions.
Thus, the key to realizing the KR map is the construction of the marginal densities ${\pi}_{\bx_{1:k}}$ in~\eqref{eq:marginals} (and hence the CDFs).

In~\cite{CuiDolgov:2021:1,CuiDolgovZahm:2023:1,ZhaoCui:2023:1}, the authors propose to construct these CDFs (and the map $\cT$) via an FTT-based \emph{approximation} to the joint density $\widehat{\pi}_{\bx} \approx \pi_{\bx}$.
To start, the square root of the unnormalized density is approximated using an FTT decomposition as described in~\cref{subsection:FTTs}, \ie, $\sqrt{p_{\bx}(\bx)} \approx \widehat{g}(\bx) \coloneqq \prod_{k=1}^{\Ntot} \bG_{k}(x_k)$.
In general, a rank-truncated TT approximation does not preserve non-negativity properties of the target function.
Constructing the TT surrogate to the square root of the target density circumvents this challenge and ensures the approximate density function $\widehat{g}\,^2$ is non-negative.
The authors of~\cite{CuiDolgov:2021:1,ZhaoCui:2023:1} further propose building a {defensive} approximation to the target density.
Given a product-form reference density $\rho_{\bx}$ satisfying $\sup_{\bx}\frac{p_\bx(\bx)}{\rho_{\bx}(\bx)} < \infty$ and a sufficiently small $\tau > 0$, the defensive FTT-based approximation to the target density is then given by
\begin{equation}
	\widehat{\pi}_{\bx}(\bx)
	=
	\frac{\widehat{p}_{\bx}(\bx)}{\widehat{Z}}
	,
	\quad
	\widehat{p}_{\bx}(\bx)
	=
	\widehat{g}(\bx)^2+\tau\rho_{\bx}(\bx)
	,
	\label{eq:FTT_defensive}
\end{equation}
where $\widehat{Z} = \int \widehat{p}_{\bx}(\bx)\, \d \bx$.
This choice of approximation ensures $\sup_{\bx} \frac{\pi_{\bx}(\bx)}{\widehat{\pi}_{\bx}(\bx)} < \infty$, which is a crucial assumption for the error estimates derived in~\cref{subsection:errors}.

In addition to preserving non-negativity of the approximate density function, building an FTT to the square root of the target function via the rank-adaptive algorithm alluded to in~\cref{subsection:FTTs} generates
\begin{equation*}
	\norm[auto]{\sqrt{p_{\bx}} - \widehat{g}}_{L^2}
	\le
	\widetilde{\varepsilon}
\end{equation*}
for some tolerance $\widetilde{\varepsilon} > 0$.
As seen in~\cref{lemma:L2toHellBound}, this bound naturally translates into a bound on the distance between the target density $\pi_{\bx}$ and its TT-based surrogate $\widehat{\pi}_{\bx}$~\eqref{eq:FTT_defensive}.
\begin{lemma}[\protect{\cite[Proposition~4 and Theorem~1]{CuiDolgov:2021:1}}]
	Assume the $L^2$ error of the FTT approximation $\widehat{g} \approx \sqrt{p_{\bx}}$ is bounded, \ie, $\norm[auto]{\sqrt{p_{\bx}} - \widehat{g}}_{L^2} \le \widetilde{\varepsilon}$.
	Furthermore, assume $\tau \le \widetilde{\varepsilon}^2$.
	Then, the Hellinger distance between $\widehat{\pi}_{\bx}$ defined in~\eqref{eq:FTT_defensive} and $\pi_{\bx}$ in~\eqref{eq:normalizedDensity} satisfies:
	\begin{equation}
		\distH{\widehat{\pi}_{\bx}}{\pi_{\bx}}
		\leq
		\varepsilon
		\coloneqq
		\frac{2 \widetilde{\varepsilon}}{\sqrt{Z}}
		.
		\label{eq:HellBound}
	\end{equation}
	\label{lemma:L2toHellBound}
\end{lemma}

As discussed above, the FTT surrogate can be constructed at a cost of $\cO\paren[auto](){\Ntot \, \Nbasis \, r^2}$ evaluations of $\sqrt{p_{\bx}}$ and $\cO\paren[auto](){\Ntot \, \Nbasis \, r^3}$ FLOPs.
The approximate conditional PDFs, $\widehat{\pi}_{x_j\given\bx_{1:j-1}} = \frac{\widehat{\pi}_{\bx_{1:j}}}{\widehat{\pi}_{\bx_{1:j-1}}}$, can then be expressed using the approximate marginal densities, which are defined as
\begin{equation}
	\widehat{\pi}_{\bx_{1:k}}(\bx_{1:k})
	=
	\frac{1}{\widehat{Z}} \paren[Big](){\tau \rho_{\bx_{1:k}}(\bx_{1:k}) + \int_{\xSpace_{>k}}{\widehat{g}(\bx)}^2 \d \bx_{k+1:\Ntot}}
	\quad
	\text{for }
	k = 1, \ldots, \Ntot-1
	,
	\label{eq:marginalFTTs}
\end{equation}
where $\rho_{\bx_{1:k}}(\bx_{1:k}) \coloneqq \prod_{i=1}^k \rho_{x_i}(x_i)$.
As detailed in~\cite[Appendix~B.3]{CuiDolgovZahm:2023:1}, the marginal densities in~\eqref{eq:marginalFTTs} can be computed analytically via a sequence of one-dimensional integrals, and the number of FLOPs required to construct these marginals, and hence the conditionals $\widehat{\pi}_{x_j \given \bx_{1:j-1}}$, is $\cO\paren[auto](){\Ntot \, \Nbasis \, r^3}$.
The cost of evaluating the corresponding inverse map $\cT$ to transform a sample $\bu \sim \unifDensity_{\bx}$ to a sample $\bx \sim \widehat{\pi}_{\bx} \approx \pi_{\bx}$ in~\eqref{eq:iRT} requires $\cO\paren[auto](){\Ntot \, \Nbasis \, r^2}$ FLOPs.

Finally, we note that the ability to construct a map~$\cS$ between any target density $\pi_{\bx}$ and the standard uniform reference~$\unifDensity_{\bx}$ is sufficient for constructing a Knothe-Rosenblatt transport between the joint density of interest~$\pi_{\bx}$ and any arbitrary product-form reference density~$\rho_{\bx}$.
More precisely, given a diagonal map $\cR$ such that $\pushforward{\cR} \rho_{\bx}= \unifDensity_{\bx}$, the composite map $\cT = \cS^{-1} \circ \cR$ defines a lower-triangular map satisfying the property $\pushforward{\cT} \rho_{\bx} = {\pi}_{\bx}$.

\subsection{Construction of \texorpdfstring{$\cT$}{transport map} for OED using tensor trains and a deep composition of KR maps}
\label{subsection:KR_OED_Difficulties}

Recall our objective to construct a transport map $\cT$ of the form~\eqref{eq:KR_map_edm} to couple a product-form reference density $\rho_{\be,\bd,\bm}$ to an approximation $\widehat{\pi}_{\be,\bd,\bm}$ of the joint density for the design, observable and inference parameter random variables.
As in~\cref{subsection:KR_OED}, the joint density $\piedm = \pi_{\bd \given \be,\bm}\pi_{\bm}\pi_{\be}$, where $\pi_{\bd \given \be,\bm} \propto \exp \paren[big](){-\frac{1}{2} \norm{\cF(\be, \bm)-\bd}_{\Gnoise(\be)^{-1}}^2}$ is the likelihood with parameter-to-observable map $\cF$, $\pi_{\bm}$ is the prior, and $\pi_{\be}$ is a user-specified density for the design parameters.
Again we let $\piedm = \frac{p_{\be,\bd,\bm}}{Z}$, assuming access only to the \emph{unnormalized} joint density, $p_{\be,\bd,\bm}$, and treating the normalizing constant $Z > 0$ as unknown.
These joint densities~$\piedm$ tend concentrate in some subdomain and exhibit complex nonlinear correlation structures, even in the relatively simple case of a linear parameter-to-observable map $\cF$, additive Gaussian noise and Gaussian prior.
\begin{example}\label{remark:toy}
	As an illustrative example, consider an inverse problem governed by Poisson's equation $-u_{xx} = m_1\sin(\frac{4}{3}x)+m_2\cos(2x)$ with constant Dirichlet data prescribed at the boundaries of the domain interval~$\Omega \coloneqq (0,2\pi)$.
	The design problem is choosing the location $e \in \Omega$ where the state $u$ should be measured to optimally infer the parameter~$\bm = [m_1,m_2]$.
	Despite the Gaussian likelihood and posterior, the joint density $\pi_{e,d}$ (visualized in~\cref{figure:ed_marginal}) is non-Gaussian even in this simple toy example.
	In particular, the density has multiple sharp peaks and concentrates along the diagonal.
	\begin{figure}[htp]
		\centering
		\includegraphics[width = \textwidth]{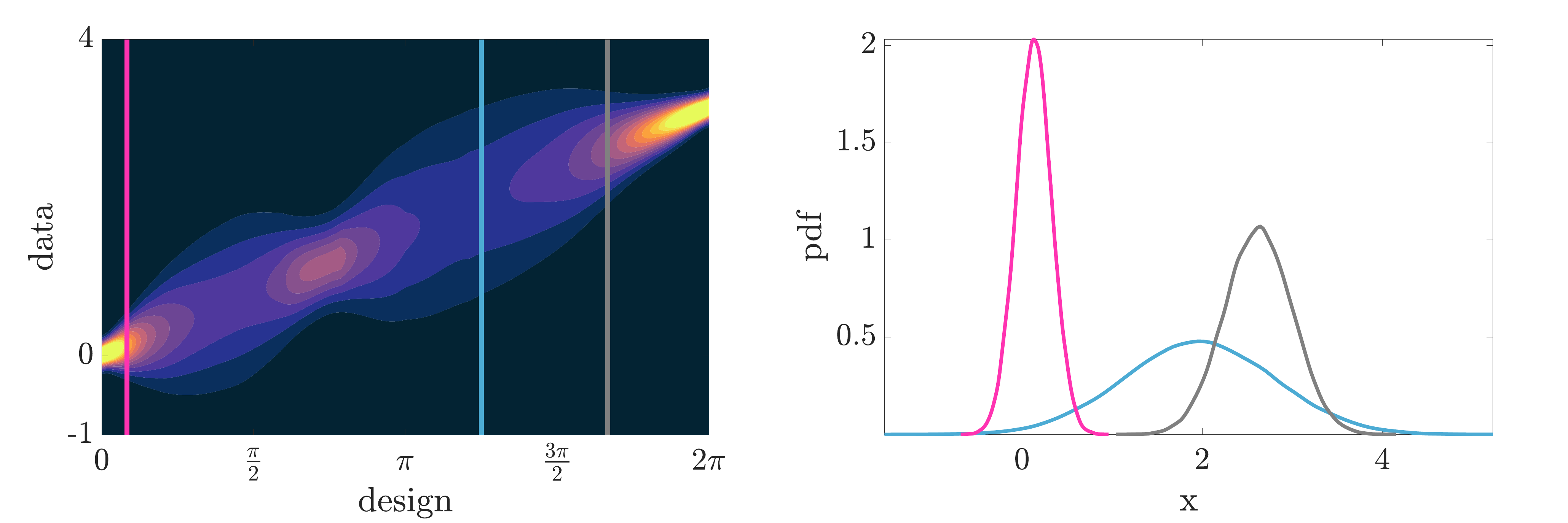}
		\caption{On the left, a contour plot of a two-dimensional density~$\pi_{e,d}$ for the illustrative toy problem in~\cref{remark:toy}.
			A visualization of~$\pi_{d \given e^*}$ for three different design locations~$e^*$ is visualized in the right image.
		The three different designs correspond to the vertical color-coded lines overlaid on the contour plot in the left.}
		\label{figure:ed_marginal}
	\end{figure}
\end{example}

In this situation, following the procedure outlined in~\cref{subsection:iRT_TT} and constructing the transport map surrogate to $\sqrt{\pedm}$ via FTT approximation in one step leads to very large ranks to ensure $\distH{\widehat{\pi}_{\be,\bd,\bm}}{\pi_{\be,\bd,\bm}} \le \varepsilon$ for suitably small error tolerances~$\varepsilon$.
Typically, the computational bottleneck of constructing the FTT decomposition for problems where the PTO map involves discretized PDE/ODE solves is the evaluation of the unnormalized density, since each evaluation requires solving the aforementioned PDE/ODE.
The computational efficiency or feasibility to construct ~$\cT$ strongly depends on the ranks, since the required number of density evaluations scales quadratically with the rank.
For certain problems this bottleneck could be mitigated by employing a reduced-order model surrogate to the parameter-to-observable map (\eg, using the discrete empirical interpolation method~\cite{ChaturantabutSorensen:2010:1}, or derivative-informed neural operators as in~\cite{WuOLearyRoseberryChenGhattas:2022:1}).\label{page:derivative-informed-neural-operators}
However, a sufficiently accurate surrogate may not be available, or may still be rather expensive to evaluate.
In general, keeping the ranks as small as possible while maintaining accuracy of the approximation is vital to the efficiency of the algorithm.

To alleviate the challenges presented by such complex densities, the authors of~\cite{CuiDolgov:2021:1} propose the \emph{deep inverse Rosenblatt transport} (DIRT), a procedure that builds a composition of transport maps guided by a suitable sequence of bridging densities.
Let $\{\pedm^\ell\}_{\ell = 1}^L$ be a sequence of unnormalized bridging densities that gradually captures the complexity of the joint target density such that $\pedm^L \coloneqq \pedm$ (see~\cref{remark:bridging} for some suitable choices).
The full composite transport map $\cT$ to approximate $\pushforward{\cT} \rhoedm = \pihatedm \approx \piedm$ is defined as
\begin{equation}
	\cT
	\coloneqq
	\cT_L
	=
	\cQ_1 \circ \cQ_2 \circ \cdots \circ \cQ_L
	.
\end{equation}
For each layer $\ell$, the composite transport map $\cT_\ell = \cQ_1 \circ \ldots \circ \cQ_\ell$ provides an approximation to the $\ell$-th normalized bridging density $\piedm^{\ell} = \frac{\pedm^{\ell}}{{Z}^{\ell}}$, \ie, $\pushforward{(\cT_\ell)} \rhoedm = \pihatedm^\ell \approx \piedm^\ell$.
Here, $Z^{\ell} > 0$ is the normalizing constant of the $\ell$-th bridging density.

The intermediate maps are constructed recursively.
At stage $\ell+1$, assuming $\cT_\ell$ is built, the next layer $\cQ_{\ell+1}$ is constructed as a squared inverse KR-map (using the procedure outlined in~\cref{subsection:iRT_TT}) satisfying $\pushforward{(\cQ_{\ell+1})} \rhoedm = q_{\be, \bd, \bm}^{\ell+1} \approx \pullback{\cT_\ell} \piedm^{\ell+1}$.
This requires building an FTT approximation to the square root of the unnormalized intermediate density $q_{\be, \bd, \bm}^{\ell+1}$.
Using~\eqref{eq:pullback_pushforward} as well as the property that $\pullback{\cT_\ell}\piedm^{\ell} \approx \rho_{\be,\bd,\bm}$, the intermediate density
\begin{equation}
	q_{\be,\bd,\bm}^{\ell+1}(\bu)
	\approx
	\pullback{\cT_\ell} \piedm^{\ell+1}(\bu)
	=
	\frac{\piedm^{\ell+1}(\cT_{\ell}(\bu))}{\piedm^{\ell}(\cT_{\ell}(\bu))} \rho_{\be,\bd,\bm}(\bu)
	\label{eq:intermediate}
\end{equation}
can be viewed as a perturbation of the reference density $\rho_{\be,\bd,\bm}$ by the ratio $\frac{\piedm^{\ell+1}(\cT_{\ell}(\bu))}{\piedm^{\ell}(\cT_{\ell}(\bu))}$.
For suitably chosen bridging densities, $q_{\be,\bd,\bm}^{\ell+1}(\bu)$ will be less complicated and concentrated than the bridging density $\piedm^{\ell+1}$, and hence easier to approximate in the FTT format.
This procedure can be seen as a successive change of coordinates or preconditioning procedure.

Since the Hellinger distance is invariant to changes in measure, the error in approximating the target $\piedm$ by the pushforward of $\rhoedm$ with respect to the final composite map $\cT_{L} = \cT_{L-1}\circ\cQ_{L}$ is
\begin{equation}
	\distH[big]{\pushforward{(\cT_{L-1}\circ\cQ_{L})}\rhoedm}{\piedm}
	=
	\distH[big]{\pushforward{(\cQ_{L})}\rhoedm}{\pullback{(\cT_{L-1})}\piedm}.
	\label{eq:lastLayerError}
\end{equation}
Thus, $\distH{\pihatedm}{\piedm}$ can be controlled by ensuring that the error in the last layer satisfies \linebreak $\distH[big]{\pushforward{(\cQ_{L})}\rhoedm}{\pullback{(\cT_{L-1})}\piedm} \le \varepsilon$ for some $\varepsilon \in (0,1]$.
This can be guaranteed by building a sufficiently accurate FTT approximation to the square root of the unnormalized pullback density \linebreak $\pullback{(\cT_{L-1})}\piedm$ as described in~\cref{lemma:L2toHellBound}.
We outline the DIRT procedure for constructing the transport map surrogate to $\piedm$ in~\cref{algorithm:DIRT}.
Once this transport map is constructed, it could be used to approximate the expected utility function exactly as described in~\cref{subsection:OED_u_KR}.
In particular, we then present the DIRT-based algorithm for computing D- and A-optimal designs in~\cref{algorithm:OED}.

\begin{algorithm}[htp]
	\caption{Deep Inverse Rosenblatt Transport (DIRT) \cite{CuiDolgov:2021:1}.}
	\label{algorithm:DIRT}
	\begin{algorithmic}[1]
		\Procedure{DIRT}{$\rho, p$}
		\Comment{Constructs a map $\cT_L$ coupling reference $\rho$ with target $\pi = \frac{p}{z}$}

		\State Choose bridging densities $\{\pedm^\ell\}_{\ell=0}^L$
		\Comment{See~\cref{remark:bridging}}
		\State $\widehat{g}^1 \gets$ FTT approx.\ to $\sqrt{p^1}$
		\State $\cT_1 \gets$ squared inverse KR map satisfying $\pushforward{(\cT_1)} \rho \approx \pi^1 = \frac{p^1}{z^1}$
		\Comment{As outlined in~\cref{subsection:iRT_TT}}
		\For{$\ell = 2, \ldots, L$}
		\State $\widehat{g}^\ell \gets$ FTT approx.\ to $\sqrt{\pullback{(\cT_{\ell-1})}p^\ell}$
		\State $\cQ_\ell \gets$ inverse KR map satisfying $\pushforward{(\cQ_{\ell})}\rho \approx \pi^{\ell} = \frac{p^{\ell}}{z^{\ell}}$
		\Comment{As outlined in~\cref{subsection:iRT_TT}}
		\State $\cT_\ell \gets \cT_{\ell-1} \circ \cQ_\ell$
		\EndFor
		\State \Return $\cT_L$
		\EndProcedure
	\end{algorithmic}
\end{algorithm}

\begin{algorithm}[htp]
	\caption{OED using the DIRT-based SAAs $\PsiX \approx \Psi_X \coloneqq \Epidge{\psi_{X}(\be, \bd)}$.}
	\label{algorithm:OED}
	\begin{algorithmic}[1]
		\Procedure{OED}{$\rhoedm, \pedm, \pi_\bm,X,N$}
		\Comment{$\pedm$: unnormalized joint density}
		\State $\cT \gets \DIRT(\rhoedm, \pedm)$
		\If{$X = D$}
		\Comment{D-optimal designs}
		\State Choose $\be^* \in \Argmax_{\be \in \eSpace} \DOpt(\rhoedm, \cT, \pi_\bm, \be,N)$
		\Comment{Using~\cref{algorithm:EIG}}
		\EndIf
		\If{$X = A$}
		\Comment{A-optimal designs}
		\State Choose number of inner samples $M$
		\State Choose $\be^* \in \Argmax_{\be \in \eSpace} \AOpt(\rhoedm, \cT, \be, N, M)$
		\Comment{Using~\cref{algorithm:AOpt}}
		\EndIf
		\State \Return $\be^*$
		\EndProcedure
	\end{algorithmic}
\end{algorithm}

\begin{remark}\label{remark:bridging}
	While~\eqref{eq:lastLayerError} shows that the approximation error of the composite map can be controlled in the last layer, independent of the choice of bridging densities,~\eqref{eq:intermediate} suggests that the choice of bridging densities plays a crucial role in the efficiency of the DIRT algorithm, since it influences the maximal ranks required for approximating $\sqrt{\pullback{(\cT_{\ell-1})}\pedm^{\ell}}$ in each layer.
	The bridging densities can be defined in various ways and what constitutes a smart choice is problem-dependent.
	One generally suitable option is the \emph{tempering} approach~\cite{GelmanMeng:1998:1}.
	Given a sequence of temperatures $0 = \beta_0 < \beta_1 < \ldots < \beta_L = 1$, the $\ell$-th tempered bridging density can be defined as $\piedm^\ell = \paren[auto](){\frac{\piedm}{\rhoedm}}^{\beta_\ell} \rhoedm$ or $\piedm^\ell = \paren[auto](){\piedm}^{\beta_\ell}$.
	Alternatively, in the case of Bayesian inference, if one is interested in approximating a highly concentrated posterior density stemming from an inverse problem with a large amount of data, it may be beneficial to build the bridging densities by adding the data sequentially in batches as $\ell$ increases.
\end{remark}

We conclude this section with some comments regarding the assumptions required in the proof of the error estimate in~\cref{theorem:ErrorBound}.
The first key assumption, $\distH{\pihatedm}{\piedm} \le \varepsilon$ for some $\varepsilon \in (0,1]$, can be guaranteed by building a sufficiently accurate FTT decomposition in the last layer of the DIRT procedure.
Secondly, the assumption of an upper bound on the ratio of the posterior density $\pi_{\bm \given \be, \bd}$ and its DIRT-based approximation $\widehat{\pi}_{\bm \given \be, \bd}$ is guaranteed by employing the defensive approximation to the target density~\eqref{eq:FTT_defensive}.

The more stringent and questionable assumption is~\eqref{eq:assumption_log}.
Since $\widehat{\pi}_{\bm \given \be, \bd} > 0$ by design, if $\pi_\bm(\bm) = 0$ for some $\bm \in \mSpace$, the assumption is clearly violated.
In practice, however, our integration domain is restricted to some bounded box that is strictly contained in the support of the prior density $\pi_\bm$.
Thus, as long as the box and prior are chosen such that $\pi_\bm$ is bounded away from zero in the restricted domain, the assumption holds, at least for Gaussian or uniform priors.
However, the bound $c_2(\be)$ from \eqref{eq:assumption_log} can get arbitrarily large as the number of experiments increases.

In practice, we find the approximation errors to be reasonable (and proportional to $\distH{\piedm}{\pihatedm}$) when the dimensions of~$\be$ and~$\bd$ are not large, \eg, when choosing a small number of experiments to perform in a batch-OED setting or a sequential OED setting, outlined in the next section.

\section{Sequential optimal experimental design using conditional DIRT}
\label{section:soed_dirt}

For certain applications, \eg medical imaging or weather prediction, data is accumulated in a sequential fashion.
With each incoming data set, the posterior distribution is updated to account for the knowledge gained from the new observations.
These settings lend themselves naturally to \emph{adaptive} or \emph{sequential} optimal design (SOED).
In SOED, experimental conditions or designs are chosen in stages to optimize an \emph{incremental} utility function that incorporates the current state of knowledge (\ie, posterior) about the unknown parameters.
In~\cref{subsection:soed}, we describe a greedy approach to solving the SOED problem and identify the main challenges in finding such optimal sequential designs.
The transport map approach for OED described in~\cref{subsection:OED_u_KR} is extended to sequential OED in~\cref{subsection:soed_tm}.
Expanding on~\cref{section:TT_construction}, we then discuss how to construct preconditioners for efficient FTT-based approximations to the SOED objective functions in~\cref{subsection:soed_preconditioners}.

\subsection{A greedy approach to sequential optimal experimental design}
\label{subsection:soed}

\begin{figure}[htp]
	\centering
	\begin{tikzpicture}
		\definecolor{mypurple}{HTML}{cc98e2}
		\tikzstyle{my box} = [fill = mypurple, rounded corners, minimum height = 6mm, minimum width = 12mm]
		\tikzstyle{abovelabel} = [midway, above = 0mm, align = center]
		\tikzstyle{belowlabel} = [midway, below = 0mm, align = center]
		\tikzstyle{centerlabel} = [midway, above = 2mm, fill = white]
		\node [my box] (design) at (-2.1,0.8) {Design};
		\node [my box] (data) at (2.1,0.8) {Data};
		\node [my box] (utility) at (0,-0.6) {Utility};

		\draw[vecArrow] (design.east) -- node [abovelabel] {experiment} (data.west);
		\draw[vecArrow] (data.south) |- node [centerlabel] {update} (utility.east);
		\draw[vecArrow] (utility.west) -| node [centerlabel] {optimize} (design.south);
	\end{tikzpicture}
	\caption{Flowchart for an SOED procedure depicting the feedback loop between finding optimal designs, using them to conduct the experiment and collect data, and updating the state of knowledge and utility function using the newly-collected data.}
	\label{figure:OED_vs_SOED}
\end{figure}
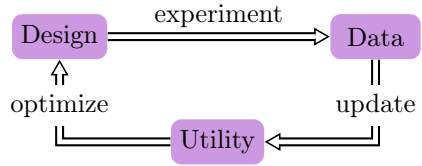

While the OED formulation described in the preceding sections focuses on finding designs given only prior information, in SOED the posterior is iteratively updated and designs are chosen adaptively in distinct stages.
As for OED, optimality is defined in terms of the level of uncertainty in the posterior distribution.
However, the key difference between OED and SOED is the inclusion of a feedback loop into the optimization problem, as visualized in~\cref{figure:OED_vs_SOED}.

For a fixed budget of $K \in \N$ experiments, the solution to the SOED problem is a set of design \emph{functions} (or policies) $\{\be_k^*(\pi_\bm^{k-1}) \}_{k=1}^{K}$ that depend on the current state of knowledge about~$\bm$, which is defined by the posterior density $\{\pi_\bm^k \}_{k=0}^{K-1}$.
Finding the optimal set of design functions requires solving a dynamic programming problem as described in~\cite[Chapter~3]{Huan:2015:1}.
However, the numerical solution of such a dynamic programming problem is extremely costly, since it involves nested optimization and thus suffers from the curse of dimensionality.

Thus, we employ a greedy approach to solving the SOED problem.
Unlike the dynamic programming formulation, the greedy formulation of SOED is myopic.
Rather than computing all $K$ optimal policies \emph{simultaneously}, future optimal experimental conditions are chosen in small batches given the current state of knowledge,  hence the curse of dimensionality is avoided.
Due to this simplification, the sequential designs $\{\be_k^*\}_{k=1}^{K}$ obtained using the greedy optimization approach will generally be \emph{sub-optimal} for a fixed number of experiments~$K$.
However, the approach does have advantages in the absence of a hard limit on the number of experiments that can be performed.
Additional experiments can be included seamlessly using the greedy approach, whereas the number of experiments must be chosen a priori in the dynamic programming context.

To facilitate the discussion, we define a recursive formula for updating the state of knowledge, or posterior density, between experimental stages:
\begin{equation}
	\pi_\bm^k
	\coloneqq
	\pi_{\bm \given \be_k, \bd_k, \bI_{k-1}}
	=
	\frac{\pi_{\bd_k \given \bm, \be_k, \bI_{k-1}} \cdot \pi^{k-1}_\bm}{\pi_{\bd_k \given \be_k, \bI_{k-1}}}
	=
	\frac{\pi_{\bd_k \given \bm, \be_k, \bI_{k-1}} \cdot \pi_{\bm \given \bI_{k-1}}}{\pi_{\bd_k \given \be_k, \bI_{k-1}}}
	,
	\label{eq:SOED_incrementalBayes}
\end{equation}
for $k = 1, \ldots, K$, where $\bI_{k-1} \coloneqq \{(\be_i^*, \bd_i^*) \}_{i=1}^{k-1}$ stores the history of optimal designs and their corresponding observations up until stage~$k-1$.
We define $\pi^{0}_\bm \coloneqq \pi_\bm$ to be the prior, and $\bI_{0}$ to be the empty set, $\bI_{0} \coloneqq \emptyset$.
In deriving the last equality, we have assumed that the stage $k-1$~posterior is independent of the decision to be made at the next stage, \ie, $\pi_{\bm \given \be_k, \bI_{k-1}} = \pi_{\bm \given \bI_{k-1}}$.

In the greedy approach to SOED, given the current state of knowledge, the $k$-th design vector~$\be_k^*$ is chosen to optimize the stage~$k$ reward,
\begin{equation}
	\Psi^k(\be_k, \pi^{k-1}_\bm)
	=
	\expect[big]{\pi_{\bd_k \given \be_k, \bI_{k-1}}}{\psi^k(\be_k, \bd_k, \pi^{k-1}_\bm)}
	,
	\label{eq:SOED_PsiK}
\end{equation}
which is an expectation of a stage~$k$ utility function~$\psi^k$.
As before, the SOED utility function is problem-specific and there are analogies to the A- and D-optimality criteria defined for OED in~\eqref{eq:A-optimality} and~\eqref{eq:D-optimality}.

For the remainder of this section, we focus on the D-optimal SOED utility function, $\psiD^k$, which for $k = 1, \ldots, K$ is defined as
\begin{equation}
	\psiD^k(\be_k, \bd_k, \pi^{k-1}_\bm)
	=
	\distKL{\pi_{\bm \given \be_k, \bd_k, \bI_{k-1}}}{\pi^{k-1}_\bm}
	=
	\distKL{\pi_{\bm \given \be_k, \bd_k, \bI_{k-1}}}{\pi_{\bm \given \bI_{k-1}}}
	.
	\label{eq:IncIG}
\end{equation}
The D-optimal, greedy sequential designs are obtained by maximizing the \emph{incremental} expected information gain in each stage of the procedure.
For generality, we allow the design and observation spaces to vary between stages and denote the stage-dependent spaces by $\eSpace_k$ and $\dSpace_k$ respectively.
The general approach for finding greedy sequential designs using the EIG criterion is summarized as follows:
\begin{enumeratearabic}
	\item \label[step]{item:greedy-approach:1}
		Initialize $\pi^{0}_\bm = \pi_\bm$, $\bI_{0} = \emptyset$, and $k = 1$.

	\item \label[step]{item:greedy-approach:2}
		At stage $k$, find an optimal $k$-th~design vector~$\be_k^*$ that maximizes
		\begin{equation*}
			\EpsiD^k(\be_k, \pi^{k-1}_\bm)
			=
			\expect[big]{\pi_{\bd_k \given \be_k, \bI_{k-1}}}{\psiD^k(\be_k, \bd_k, \pi^{k-1}_\bm)}
		\end{equation*}
		over $\be_k \in \eSpace_k$.

	\item \label[step]{item:greedy-approach:3}
		Perform experiment~$k$ with design vector~$\be_k^*$ to obtain data~$\bd_k^* \in \dSpace$.

	\item \label[step]{item:greedy-approach:4}
		Update $\bI_k = \{\bI_{k-1}, \be_k^*, \bd_k^* \}$ and $\pi^k_\bm = \pi_{\bm \given \bI_k}$.

	\item \label[step]{item:greedy-approach:5}
		Increment~$k$ and repeat \cref{item:greedy-approach:2,item:greedy-approach:3,item:greedy-approach:4} until all experiments have been performed or sufficient reduction in posterior uncertainty has been achieved.
\end{enumeratearabic}
In the following section we focus on \cref{item:greedy-approach:2} and extend the KR-based approach of~\cref{section:transport_OED} to numerically approximate the incremental expected information gain~$\EpsiD^k$.

\subsection{Knothe-Rosenblatt-based approximation to sequential D-optimal BOED}
\label{subsection:soed_tm}

Assume we have access to a KR map $\cT^k$ coupling a reference density $\rhoedm$ to the $k$-th joint conditional density, which in the most general form satisfies
\begin{equation}
	\pi_{\be_k, \bd_k, \bm \given \bI_{k-1}}
	=
	\pi_{\bd_k \given \be_k, \bm, \bI_{k-1}} \cdot \pi_{\bm \given \bI_{k-1}} \cdot \pi_{\be_k \given \bI_{k-1}}
	= \pi_{\bd_k \given \be_k, \bm, \bI_{k-1}} \cdot \pi_{\bm}^{k-1} \cdot \pi_{\be_k \given \bI_{k-1}}
	.
	\label{eq:kthdensity}
\end{equation}
The $k$-th incremental likelihood $\pi_{\bd_k \given \be_k, \bm, \bI_{k-1}}$ and the design density $\pi_{\be_k \given \bI_{k-1}}$ are both specified via the model (see~\cref{remark:incrementalLikelihood}).
Then the stage~$k$ incremental expected information gain,
\begin{equation*}
	\Psi_{D}^k(\be_k, \pi^{k-1}_\bm)
	=
	\expect[big]{\pi_{\bd_k \given \be_k, \bI_{k-1}}}{\psi^k_D(\be_k, \bd_k, \pi^{k-1}_\bm)}
	,
\end{equation*}
can be approximated at any $\be_k\in\eSpace_k$ using~\cref{algorithm:EIG},
\begin{equation}
	\label{eq:EIG}
	\Psi^k_D(\be_k, \pi_{\bm}^{k-1})
	\approx
	\widehat{\Psi}^k_D(\be_k,\pi_{\bm}^{k-1})
	\coloneqq
	\textsc{DOpt}(\rhoedm, \cT^k, \pi_{\bm}^{k-1}, \be_k, N)
	.
\end{equation}

While this approach is straightforward, it suffers from a few drawbacks.
The first major issue is that we generally do not have a closed-form expression for the posterior $\pi_{\bm}^{k-1}$, hence we cannot execute step 6 of~\cref{algorithm:EIG}, which requires evaluation of the density $\pi_{\bm}^{k-1}$ at the sample parameters.
The second major issue is a computational one.
While we could employ~\cref{algorithm:DIRT} to construct an approximate map $\cT^k$ at each stage, the target density $\pi_{\be_k,\bd_k,\bm\given \bI_{k-1}}$ gets progressively more concentrated as data is collected.
Thus, constructing the KR rearrangement using DIRT gets successively more costly in terms of the number of unnormalized density evaluations and FLOPs, as $k$ increases.
A sufficiently accurate approximation also requires an increasing number of layers in DIRT.
Additionally, the posteriors tend to concentrate sequentially to some subdomain as data is collected and we should exploit this structure to avoid wasting computational resources.
All those challenges are addressed in \cref{subsection:soed_preconditioners}.

\begin{remark}[Incremental likelihood and density]
	The exact forms of the incremental likelihood \linebreak $\pi_{\bd_k \given \be_k, \bm, \bI_{k-1}}$ and the $k$-th~design density $\pi_{\be_k \given \bI_{k-1}}$ are model-dependent.
	For example, if one couples the $k$-th~observations $\bd_k \in \dSpace_k$, $k$-th~designs $\be_k \in \eSpace_k$, and the inference parameters~$\bm$ via
	\begin{equation}
		\bd_k
		=
		\cF(\be_k, \bm)
		+
		\boldeta_k
		,
		\quad
		\boldeta_k
		\sim
		\normal{0}{\sigma_k^2 \, \id}
		,
	\end{equation}
	and assumes the only explicit dependence on the history vector $\bI_{k-1}$ is through the prior $\bm \sim \pi_{\bm \given \bI_{k-1}}$ at stage $k$, then the stage-$k$ likelihood is independent of $\bI_{k-1}$ and satisfies
	\begin{equation}
		\pi_{\bd_k \given \be_k, \bm}
		\propto
		\exp \paren[Big][]{-\frac{1}{2\sigma_k^2} \norm{\cF(\be_k, \bm) - \bd_k}^2}
		.
	\end{equation}
	This is the form of the $k$-th likelihood used in~\cref{section:NumericalExamples}.
	Likewise, the design density is independent of $\bI_{k-1}$, \ie, $\pi_{\be_k \given \bI_{k-1}} = \pi_{\be_k}$.
	\label{remark:incrementalLikelihood}
\end{remark}

\subsection{Preconditioned Knothe-Rosenblatt maps for SOED}
\label{subsection:soed_preconditioners}

To alleviate the drawbacks of the straightforward approach, we propose a preconditioning method that enables incorporation of previously acquired knowledge.
This approach helps reduce the number of layers required in the sequential construction of the DIRT maps $\{\cT^k \}_{k=1}^{K}$ and in the amount of unnormalized density evaluations needed for constructing the FTT decomposition.
The idea driving the preconditioning approach is similar to the reasoning behind the layered construction of deep inverse Rosenblatt transports in~\cite{CuiDolgov:2021:1}.

Assume we have access to a KR-based approximation of the posterior, $\pushforward{(\cT_{\bm}^k)}\rho_{\bm} = \widehat{\pi}^k_\bm \approx \pi_{\bm \given \bI_k} \eqqcolon \pi_\bm^k$ for $k = 1, \ldots, K-1$.
To start, we approximate the $k$-th joint conditional density in~\eqref{eq:kthdensity} using the corresponding transport map surrogate,
\begin{equation}
	\pi_{\be_k,\bd_k,\bm \given \bI_{k-1}} \approx \pi_{\bd_k \given \be_{k},\bm,\bI_{k-1}} \cdot \widehat{\pi}_{\bm \given \bI_{k-1}} \cdot \pi_{\be_{k} \given \bI_{k-1}} \eqqcolon \widetilde{\pi}_{\be_{k},\bd_{k},\bm \given \bI_{k-1}}.
\end{equation}
Then, we construct a KR rearrangement to the pullback density, $\cK^k$ to $\pullback{(\cL^k)}\widetilde{\pi}_{\be_{k},\bd_{k},\bm \given \bI_{k-1}}$,
where $\cL^k$ is a stage~$k$ diagonal preconditioner of the form
\begin{equation}
	\cL^k(\bz)
	\coloneqq
	\cL^k({\bz_e},{\bz_d},{\bz_m})
	=
	\begin{bmatrix*}[l]
		\cR^k_\be({\bz_e})
		\\
		\id_\bd({\bz_d})
		\\
		\cT^k_\bm(\bz_m)
	\end{bmatrix*}
	\label{eq:KR_precondPost}
\end{equation}
with ${\bz_e} \in \R^\Ne$, ${\bz_d} \in \R^{\Nd}$, ${\bz_m} \in \R^\Nm$ and $\cR^k_\be$ denoting a KR rearrangement such that $\pushforward{(\cR^k_\be)} \rho_\be= \pi_{\be_k \given \bI_{k-1}}$.
If $\cR^k_\be$ is not known, it could be replaced by the identity operator $\id_\be$.
Using the preconditioner~\eqref{eq:KR_precondPost} such that $\pushforward{(\cL^k)}\rhoedm \propto \pi_{\be_k}\rho_{\bd}\widehat{\pi}_{\bm \given \bI_{k-1}}$, the $k$-th~preconditioned density $\pullback{(\cL^k)}\widetilde{\pi}_{\be_{k},\bd_{k},\bm \given \bI_{k-1}}$, which we aim to approximate, is
\begin{equation*}
	\pullback{(\cL^k)}\widetilde{\pi}_{\be_{k},\bd_{k},\bm \given \bI_{k-1}}
	\propto
	\pi_{\bd_k \given \be_k, \bm, \bI_{k-1}} \paren[auto](){{\bz_d} \given \cR^k({\bz_e}), \cT^k_\bm(\bz_m)} \rho_{\be}(\bz_e)\rho_{\bm}(\bz_m).
\end{equation*}
The KR map $\cK^k$ is built using~\cref{algorithm:DIRT}, such that $\pushforward{(\cK^k)}\rhoedm \approx \pullback{(\cL^k)}\widetilde{\pi}_{\be_{k},\bd_{k},\bm \given \bI_{k-1}}$.
This requires significantly fewer density evaluations in practice than constructing a KR rearrangement to $\pi_{\be_{k},\bd_{k},\bm\given\bI_{k-1}}$.
Once $\cK^k$ is constructed, the composite map $\cT^k = \cL^k \circ \cK^k$ provides a coupling between the reference $\rhoedm$ and the $k$-th approximation to the joint target $\widetilde{\pi}_{\be_k,\bd_k,\bm \given \bI_{k-1}}$.

The maps $\cT^k_{\bm}$ required for defining the preconditioner are constructed in a recursive fashion using a running approximation to the posterior density $\pi_{\bm \given \bI_{k-1}}$.
Assuming $\pushforward{(\cT^{k-1}_{\bm})}\rho_{\bm} \approx \pi_{\bm \given \bI_{k-1}}$, $\cT^k_{\bm} \coloneqq \cT^{k-1}_{\bm} \circ \cK^k_{\bm}$, where $\cK_{\bm}^k$ is a KR map satisfying $\pushforward{(\cK_{\bm}^k)}\rho_{\bm} = \pullback{(\cT^{k-1}_{\bm})}\pi_{\bm \given \bI_{k}}$.
This could be viewed as employing a DIRT-based construction to the stage $K-1$ posterior $\pi_{\bm \given \bI_{K-1}}$ guided by the bridging densities $\{\pi_{\bm\given\bI_{k}}\}_{k=1}^{K-1}$.
As mentioned in~\cref{remark:bridging}, these bridging densities are a natural choice for posteriors stemming from Bayesian inverse problems with large amounts of data, or for sequential Bayesian inference.

\begin{remark}[DIRT map approximation to the posterior densities]
	\label{remark:DIRT_posterior}
	Since the sequence of posterior densities $\{\pi_{\bm \given \bI_k}\}_{k=1}^{K-1}$ are obtained by performing experiments with conditions that maximize the incremental expected information gain $\distKL{\pi_{\bm \given \be_{k+1}, \bd_{k+1}, \bI_k}}{\pi_{\bm \given \bI_k}}$, the distance between adjacent posteriors may be quite large.
	Thus, the ranks required to approximate the square root of $\pullback{(\cT^{k-1}_{\bm})}\pi_{\bm \given \bI_{k}}$ in one step may be large.
	For certain problems it is more efficient to combine the tempering approach discussed in~\cref{subsection:KR_OED_Difficulties} with the sequential data accumulation approach discussed in~\cref{subsection:soed_preconditioners} to obtain a composite set of bridging densities.
\end{remark}

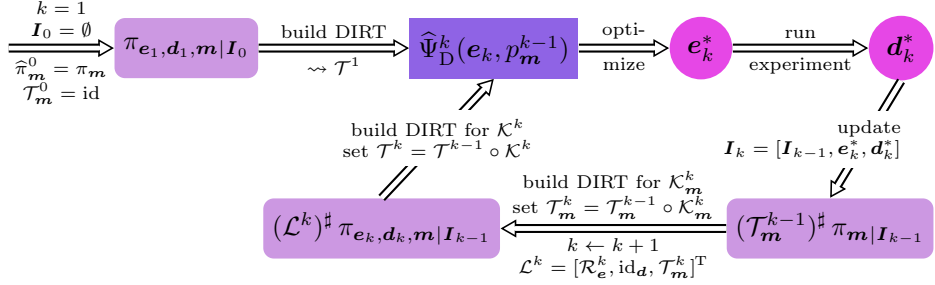
\begin{figure}[htp]
	\centering
	\begin{tikzpicture}
		\definecolor{mypurple}{HTML}{cc98e2}
		\definecolor{mypurple2}{HTML}{8f63e5}
		\definecolor{mypurple3}{HTML}{e647e6}
		\tikzstyle{my rounded box} = [fill = mypurple, rounded corners, minimum height = 8mm, minimum width = 10mm]
		\tikzstyle{my box} = [fill = mypurple2, minimum height = 8mm, minimum width = 10mm]
		\tikzstyle{my circle} = [circle, fill = mypurple3, minimum height = 8mm]
		\tikzstyle{abovelabel} = [midway, above = 0mm, align = center, font = \scriptsize]
		\tikzstyle{belowlabel} = [midway, below = 0mm, align = center, font = \scriptsize]
		\tikzstyle{centerlabel} = [midway, align = center, fill = white, font = \scriptsize]
		\node [my rounded box] (one) at (0.0,2.0) {$\pi_{\be_1, \bd_1, \bm \given \bI_{0}}$};
		\node [my box, right = 20mm of one] (two) {$\EhatPsiD^k(\be_k, p_\bm^{k-1})$};
		\node [my circle, right = 12mm of two] (three) {$\be_k^*$};
		\node [my circle, right = 18mm of three] (four) {$\bd_k^*$};
		\node [my rounded box, below = 24mm of four.east, anchor = east] (five) {$\pullback{(\cT^{k-1}_\bm)} \pi_{\bm \given \bI_{k-1}}$};
		\node [my rounded box, left = 30mm of five] (six) {$\pullback{(\cL^k)} \pi_{\be_k, \bd_k, \bm \given \bI_{k-1}}$};
		\node [left = 14mm of one] (zero) {};

		\draw[vecArrow] (zero.east) -- node [abovelabel] {$k = 1$ \\ $\bI_{0} = \emptyset$} node [belowlabel] {$\widehat{\pi}_\bm^{0} = \pi_\bm$ \\ $\cT_\bm^{0} = \id$} (one.west);
		\draw[vecArrow] (one.east) -- node [abovelabel] {build DIRT} node [belowlabel] {{$\leadsto \cT^1$}} (two.west);
		\draw[vecArrow] (two.east) -- node [abovelabel] {opti-} node [belowlabel] {mize} (three.west);
		\draw[vecArrow] (three.east) -- node [abovelabel] {run} node [belowlabel] {experiment} (four.west);
		\draw[vecArrow] (four.south) -- node [centerlabel, left = -6mm, align = right] {update \\ $\bI_k = [\bI_{k-1}, \be_k^*, \bd_k^*]$} (five.north);
		\draw[vecArrow] (five.west) -- node [abovelabel] {build DIRT for $\cK_\bm^k$ \\ set $\cT_\bm^k = \cT_\bm^{k-1} \circ \cK_\bm^k$} node [belowlabel] {$k \gets k+1$ \\ $\cL^k = [\cR_\be^k, \id_\bd, \cT^k_\bm]^\transp$} (six.east);
		\draw[vecArrow] (six.north) -- node [centerlabel] {build DIRT for $\cK^k$ \\ set $\cT^k = \cT^{k-1} \circ \cK^k$} (two.south);
	\end{tikzpicture}
	\caption{Flowchart visualization of the greedy preconditioned SOED procedure outlined in~\cref{algorithm:SOED_EIG}.}
	\label{figure:OED_vs_SOEDv3}
\end{figure}

Due to the recursive layered approximation to the posterior density, there is a sequential accumulation of error in the KR approximation to $\pi_{\be_k, \bd_k, \bm \given \bI_{k-1}}$.
However, this error can be controlled using~\cref{lemma:L2toHellBound}.
The amount of additional applications of the PTO map required to keep the running approximation to the posterior density is generally offset by the reduction in computational work due to using the resulting preconditioning operator $\cL^k$ defined in~\eqref{eq:KR_precondPost}.
We present the full posterior-preconditioned SOED procedure in~\cref{algorithm:SOED_EIG}, where we again leave the optimization algorithm for maximizing the incremental EIG arbitrary.
A visual representation of the algorithm in the form of a flowchart is also included in~\cref{figure:OED_vs_SOEDv3}.

\begin{algorithm}[htp]
	\caption{Greedy posterior-preconditioned SOED algorithm using the incremental EIG criterion.}
	\label{algorithm:SOED_EIG}
	\begin{algorithmic}[1]
		\Procedure{SOED}{$K, \rhoedm, \pi_{\be_1}, \pi_\bm, \pi_{\bd_1 \given \be_1, \bm}$}
		\State $\cL^0 \gets \id_{\Ntot}$, $\cT^{0}_\bm \gets \id_{\Nm}$, $\widehat{\pi}_\bm^{0} \gets \pi_\bm$, $\bI_{0} = \emptyset$
		\Comment{Initialization}
		\For{$k = 1, \ldots, K$}
		\State $\pi_{\be_k, \bd_k, \bm \given \bI_{k-1}} \gets \pi_{\bd_k \given \be_k, \bm, \bI_{k-1}} \cdot \widehat{\pi}_\bm^{k-1} \cdot \pi_{\be_k \given \bI_{k-1}}$
		\State $\cK^k \gets \DIRT(\rhoedm, \pullback{(\cL^k)} \pi_{\be_k, \bd_k, \bm \given \bI_{k-1}})$
		\Comment{Using~\cref{algorithm:DIRT}}
		\State $\cT^k \gets \cL^k \circ \cK^k$
		\State Choose $\be_k^* \in \Argmax_{\be_k \in \eSpace} \DOpt(\rhoedm, \cT^k,\widehat{\pi}_\bm^{k-1}, \be_k)$
		\Comment{Using~\cref{algorithm:EIG}}
		\State Perform experiment with $\be_k^*$ and acquire $\bd_k^*$
		\State Update $\bI_k = \{\bI_{k-1}, \be_k^*, \bd_k^*\}$
		\State $\cK^k_\bm \gets \DIRT(\rho_\bm, \pullback{(\cT_\bm^{k-1})} \pi_{\bm \given \bI_k})$
		\Comment{Using~\cref{algorithm:DIRT}}
		\State $\cT^k_\bm = \cT^{k-1}_\bm \circ \cK^k_\bm$, $\widehat{\pi}^{k}_{\bm} \gets \pushforward{(\cT^k_\bm )}\rho_{\bm}$
		\State Update $\cL^k$ using equation~\eqref{eq:KR_precondPost} and $\cT^k_\bm$
		\EndFor
		\State \Return $\{\be_k^*\}_{k=0}^{K-1}$, $\{\bd_k^*\}_{k=0}^{K-1}$, $\cT^{K-1}_\bm$
		\EndProcedure
	\end{algorithmic}
\end{algorithm}

\section{Numerical examples}
\label{section:NumericalExamples}

We first illustrate our approach and compare it to nested Monte Carlo approximation for a simple nonlinear model in~\cref{subsection:nonlin_comparison}.
We then demonstrate the effectiveness of our proposed method on finding optimal designs for two model problems.
In~\cref{subsection:SEIR_example}, we consider a non-intrusive sensor placement problem using transport-map-based procedures (\cref{algorithm:SOED_EIG} and \cref{algorithm:OED}) to compute optimal testing times for an inverse problem in disease modeling.
In~\cref{subsection:Poisson_example}, we consider instead intrusive designs, \ie, we choose optimal boundary conditions for an inverse problem with an elliptic forward operator.
The algorithms used in both examples are implemented in \matlab using the DIRT~FTT codes from the {\texttt{deep-tensor toolbox}}~\cite{Cui:DIRT}.

\subsection{Simple nonlinear model}
\label{subsection:nonlin_comparison}

To provide some intuition, we first consider the simple nonlinear algebraic model from~\cite[Section~5]{HuanMarzouk:2013:1}.
The design-dependent model relating the design $e \in [0,1]$ and scalar parameter~$m$ to observation~$d$ is
\begin{equation}
	d(e,m)
	=
	e^2 \, m^3 + m \, \exp(-\abs{0.2-e}) + \eta
	,
	\label{eq:nonlinFwd}
\end{equation}
where $m$ has a uniform prior on $[0,1]$, and $\eta \sim \cN(0,10^{-4})$ is normally distributed.

This model is inexpensive to evaluate and possesses key features of our target applications: it is nonlinear in both the design and parameters, and it utilizes a non-Gaussian prior.
The experimental goal is to choose designs that maximize the expected information gain $\EpsiD$~\eqref{eq:D-objectiveSAA} from prior to posterior.
For the following simulations, we compute these designs in two ways:
\begin{enumerate*}[label=\ensuremath{(\roman*)}]
	\item
		using our proposed transport map approach (yielding an optimal design $e_{\textup{OED-DIRT}}^*$), and,

	\item
		using a nested (or double-loop) Monte Carlo (NMC) method~\cite{HuanMarzouk:2013:1}, combined with a greedy optimization approach (yielding an optimal design $e_{\textup{OED-NMC}}^*$).
\end{enumerate*}

For our DIRT-based approach, we use~\cref{algorithm:OED} with $X = D$ and $N = \num{1000}$ samples for the MC estimator, and \matlab's \texttt{fmincon} to optimize the EIG.
For the NMC approach, the EIG is approximated as
\begin{equation}
	\EpsiNMC(e)
	=
	\frac{1}{N_{\textup{out}}} \sum_{i=1}^{N_{\textup{out}}} \log \paren[big](){\pi_{d \given e, m}(d^{(i)}\given e, m^{(i)})}
	-
	\log \paren[big](){\widehat{\pi}_{d \given e }(d^{(i)}\given e)}
	,
	\label{eq:NMC}
\end{equation}
where $m^{(i)}$ are drawn from the prior, and $d^{(i)}$ is drawn from the resulting conditional distribution of $d \given e,m^{(i)}$.
The evidence is approximated by $\widehat{\pi}_{d \given e}(d^{(i)} \given e) = \frac{1}{N_{\textup{in}}} \sum_{j=1}^{N_{\textup{in}}} \pi_{d \given e,m}(d^{(i)} \given e, m^{(i,j)})$, where $m^{(\cdot,j)}$ are samples from the prior.
For our simulations, we set $N \coloneqq N_{\textup{out}} = N_{\textup{in}}$ and reuse the prior samples $m^{(i)}$ drawn for the outer loop to evaluate the inner loop.
As described in~\cite[Section~2.3]{HuanMarzouk:2013:1}, while sample reuse contributes to estimator bias, it significantly reduces the cost of evaluating $\EpsiNMC$ at any design from the product $N_{\textup{in}} N_{\textup{out}}$ to $N_{\textup{out}}$ model evaluations.

A common approach to computing optimal designs is the so-called greedy approach~\cite{WuOLearyRoseberryChenGhattas:2022:1,WuChenGhattas:2023:1,AlexanderianNicholsonPetra:2022:1}.
The greedy approach is easy to implement as it does not require derivatives of the objective function.
While this approach is typically used for sensor selection problems, extending it to our model problem is straightforward.
To this end, we discretize our design domain $[0,1]$ into $51$ equally-spaced candidate design nodes.
Computing $k = 1,2$ designs that maximize the EIG using the greedy approach thus amounts to $51$ and $102$ evaluations of the nested MC estimator $\EpsiNMC$, respectively.

\begin{figure}
	\centering
	\begin{subfigure}{0.7\textwidth}
		\centering
		\includegraphics[width = \textwidth]{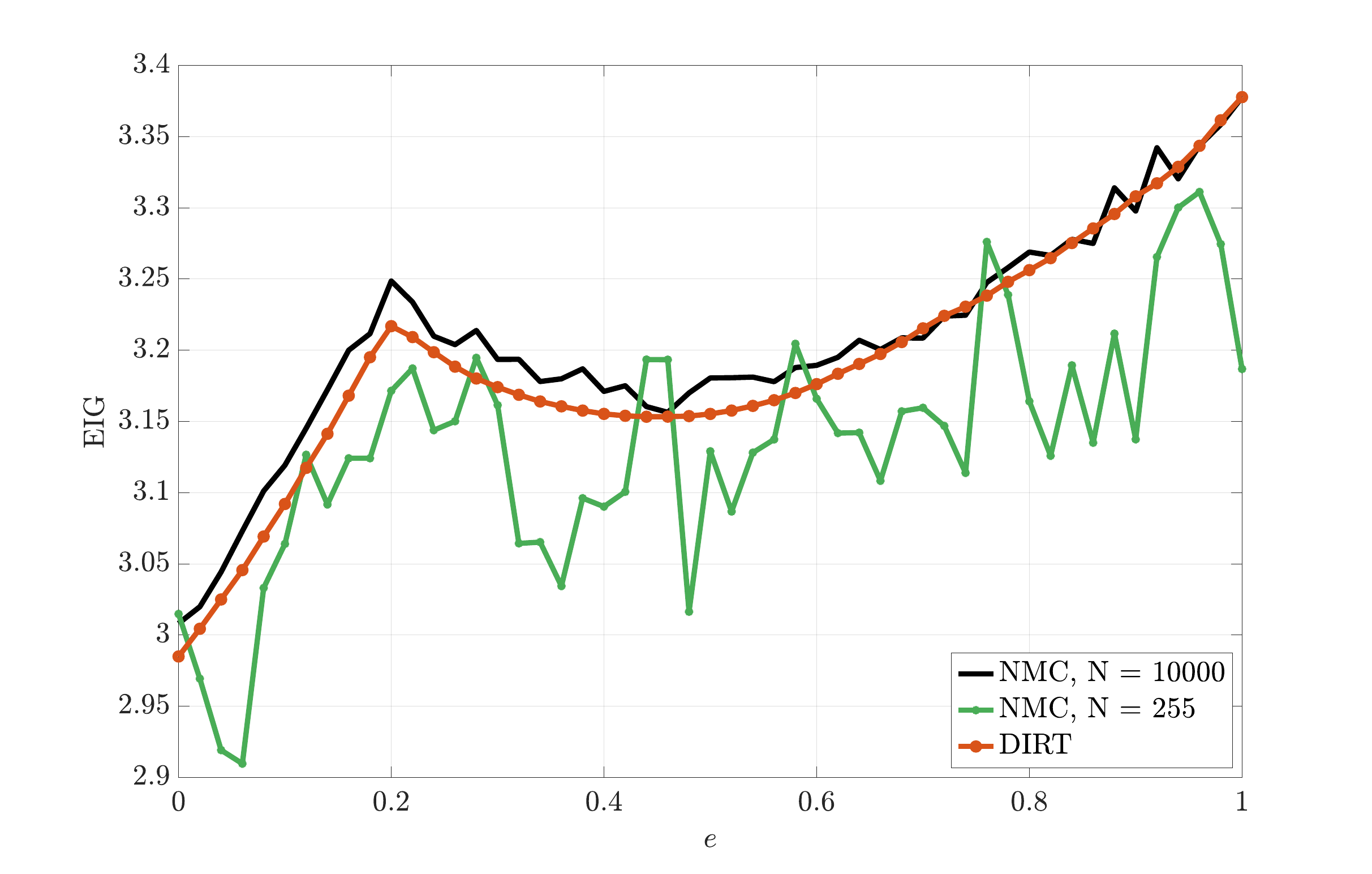}
	\end{subfigure}%
	\begin{subfigure}{0.3\textwidth}
		\centering
		\includegraphics[width = 0.75\textwidth]{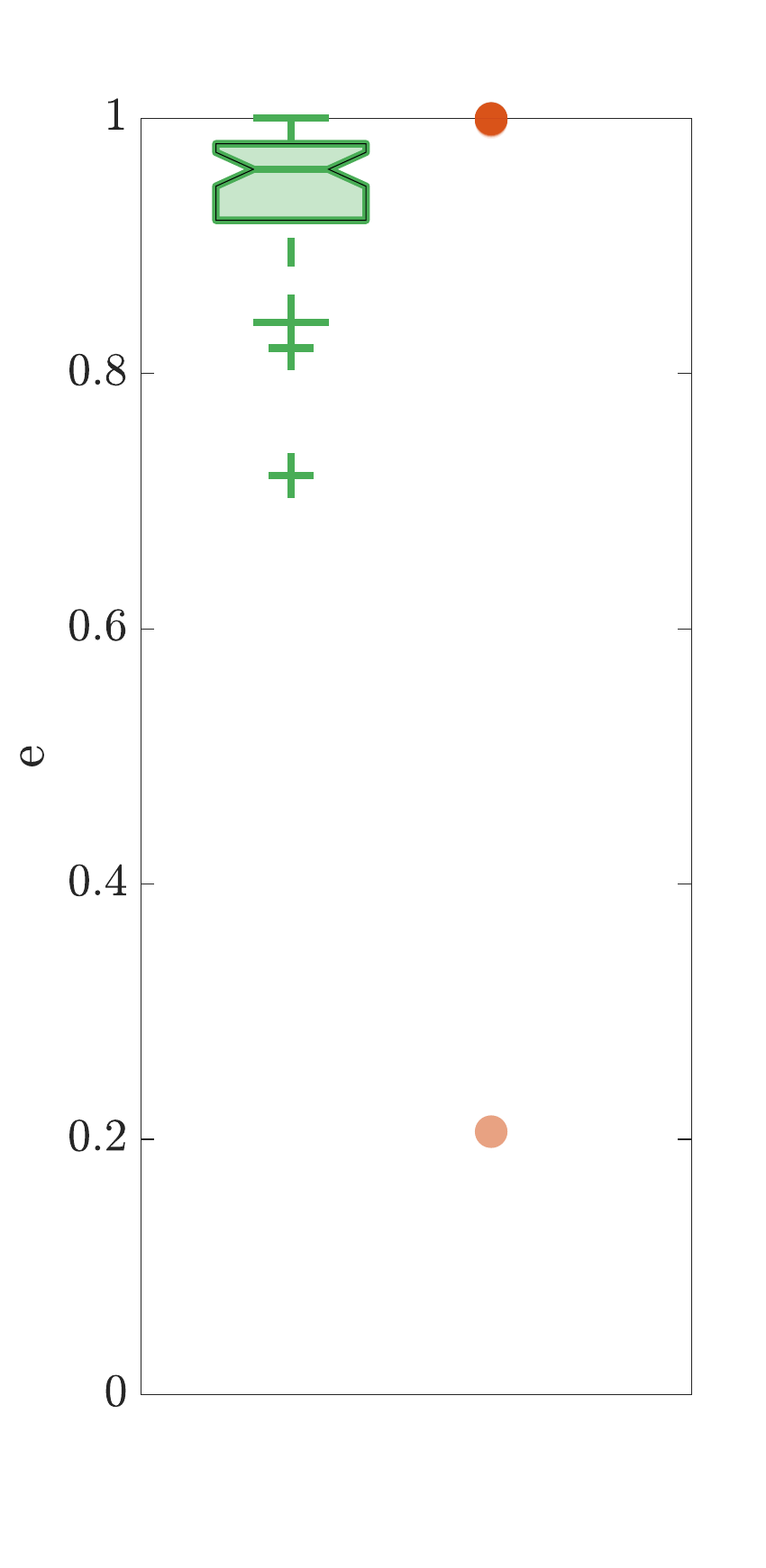}
	\end{subfigure}
	\caption{The expected information gain at different designs for the nonlinear example in~\cref{subsection:nonlin_comparison} (left), and a comparison of the spread of optimal designs obtained using NMC with $N = 255$ and our approach (right). A scatter plot is used to show the optimal designs obtained using our approach, whereas a boxplot is used for the NCM optimizers.}
	\label{fig:nonlin_1design_EIG}
\end{figure}

\textbf{One design case:} In the case of designing one experiment, computing $e_{\textup{OED-DIRT}}^*$ required constructing a transport map approximation to the joint density on $3$~variables ($e,d,m$).
This construction required \num{12960} evaluations of the forward model~\eqref{eq:nonlinFwd}.
To make the comparison with NMC fair, we fix the total model evaluation budget to \num{12960}.
When using the greedy approach, this amounts to setting $N \approx 255$.
A visual comparison of the approximate EIG at each candidate design, evaluated using our approach, and nested MC with $N = 255$ (as well as $N = \num{10000}$) is depicted in~\cref{fig:nonlin_1design_EIG}.
Since the Monte Carlo approximation of the EIG depends on the samples used, we compute the optimal designs \num{50}~times using both approaches.
The distribution of optimal designs computed using both approaches is visualized in~\cref{fig:nonlin_1design_EIG}.
The optimal designs obtained using NMC are spread out over $[0.72,1]$ (due to the small number of samples used), whereas the optimal design found using~\cref{algorithm:OED} are primarily concentrated around the true optimizer, $e = 1$.
The kink at $e = 0.2$ is sometimes found instead of the global maximizer (in this particular case, $6$ out of $50$~runs).
Using a larger number of samples for the NMC estimator would of course decrease the spread of $e_{\textup{OED-NMC}}^*$.
However, we emphasize that if a more reasonable number of samples were used for the NMC estimator, \eg, $N = \num{10000}$ (which produces the black curve in~\cref{fig:nonlin_1design_EIG}), even two evaluations of $\EpsiNMC$ would require more forward solves than our approach.

\begin{figure}
	\centering
	\includegraphics[width = 0.95\textwidth]{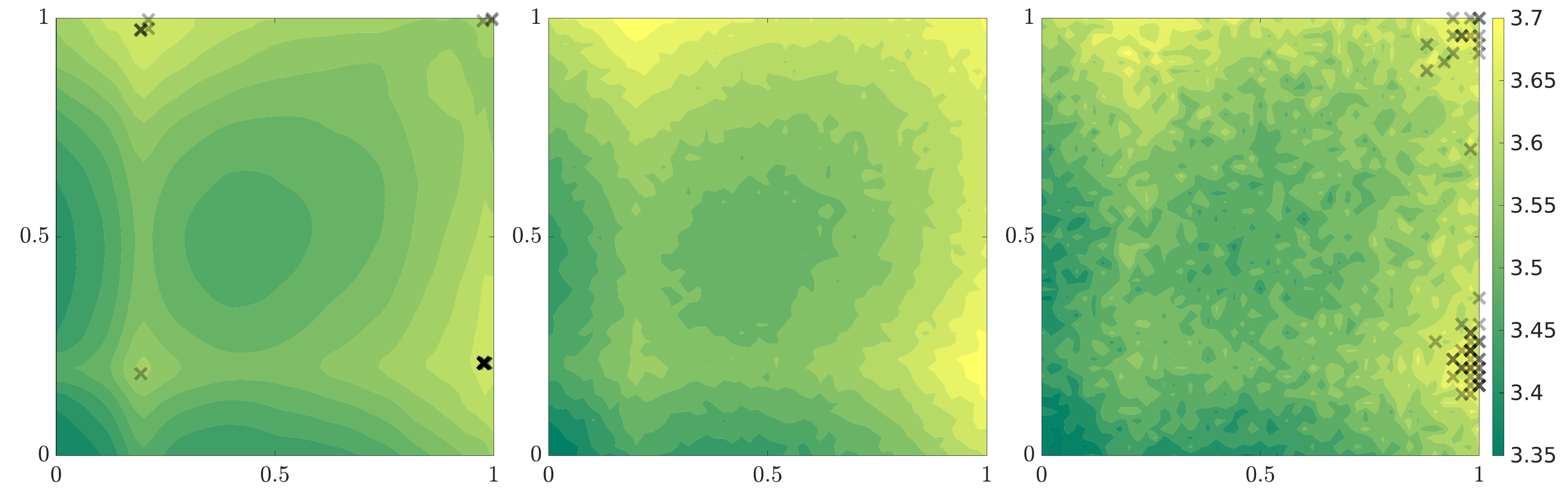}
	\caption{The expected information gain for the two design nonlinear example in~\cref{subsection:nonlin_comparison}. On the left is the EIG as approximated using our approach, on the right is the EIG approximated using NMC with $N = 828$, and the middle is the EIG approximated with NMC using $N = \num{10000}$ samples per design point. The optimal designs obtained using both approaches are visualized using the \textbf{x} symbol in the left and right plots.}
	\label{fig:nonlin_2design_EIG}
\end{figure}

\textbf{Two design case:} In the case of designing two experiments, computing $\be_{\textup{OED-DIRT}}^*$ required constructing a transport map approximation to a joint density of $5$~variables, which took \num{84448} evaluations of the forward model.
For the NMC approach, we again fix the total number of forward model evaluations to this number, which amounts to setting $N = 828$ for the greedy approach.
The global maximizers of the EIG in this case are at $\be^* = [0.2,1]^\transp$ and $\be^* = [1,0.2]^\transp$, the latter of which is attainable using the greedy approach.
~\Cref{fig:nonlin_2design_EIG} visualizes the (approximate) EIG for the two design case using our approach, as well as NMC with $N = 828$ and $N = \num{10000}$ (as a reference).
Once again, we run each optimization algorithm $50$~times and plot the optimal designs in~\cref{fig:nonlin_2design_EIG}.
As in the one design case, the designs obtained via the greedy approach and the NMC estimator are more widespread.
For our approach, \num{46} of the 50 optimizers are clustered around the global optimal design locations $[0.2,1]^\transp$ and $[1,0.2]^\transp$, three get stuck close to the local maximum at $[1,1]^\transp$, and one gets stuck at another local maximum at $[0.2,0.2]^\transp$.
Again, using NMC with a larger number of samples would reduce the spread of the optimal designs.
In particular, setting $N = \num{10000}$ produces reasonable estimates of the EIG (as visualized in the middle of \cref{fig:nonlin_2design_EIG}), however in this case, $9$ evaluations of the estimator would already require more forward model solves than our approach.

These results suggest that our method is competitive with the nested Monte Carlo approach for batch OED involving a small number of experiments.
Additionally, as described in~\cref{section:soed_dirt}, our approach is easily adapted to sequential experimental design, which is a challenging problem for conventional methods.
The nested MC approximator to the EIG could be adapted to approximate the incremental EIG~\eqref{eq:IncIG}, but without additional modifications, evaluating the approximate EIG after the first stage would require drawing samples from an intractable posterior distribution.
In the following examples, we illustrate the effectiveness of our approach for sequential OED and for larger-dimensional problems.

\subsection{Optimal observation times for parameter inversion using the SEIR disease model}
\label{subsection:SEIR_example}

Next, we consider the susceptible-exposed-infected-removed (SEIR) model, commonly used for describing the spread of infectious diseases.
Under the assumption of constant population size, the SEIR model is comprised of the following system of ordinary differential equations,
\begin{equation}
	\begin{aligned}
		\frac{\d S}{\d t}
		&
		=
		- \beta(t) \, S \, I
		,
		&
		&
		\frac{\d E}{\d t}
		&
		=
		\beta(t) \, S \, I - \alpha E
		,
		\\
		\frac{\d I}{\d t}
		&
		=
		\alpha E - \gamma(t) \, I
		,
		&
		&
		\frac{\d R}{\d t}
		&
		=
		(\gamma^r + \gamma^d(t)) \, I
		,
	\end{aligned}
	\label{eq:SEIR}
\end{equation}
where the variables $S(t), E(t), I(t), R(t)$ are used to denote the fractions of susceptible, exposed, infected and removed individuals at time~$t$, respectively, and are initialized with $S(0) = 99$, $E(0) = 1$, and $I(0) = R(0) = 0$.
The parameters to be estimated are $\beta(t), \alpha, \gamma^r, \gamma^d(t)$, where the constants $\alpha$ and $\gamma^r$ denote the rate of susceptibility to exposure and infection to recovery, respectively.
To simulate the effect of policy changes or other time-dependent factors (\eg, quarantine and overcrowding of hospitals), the rates at which exposed individuals become infected and infected individuals perish are assumed to be time-dependent and parametrized as follows:
\begin{equation*}
	\beta(t)
	=
	\beta_1 + \frac{\tanh(7(t-\tau))}{2}(\beta_2-\beta_1)
	,
	\quad
	\gamma^d(t)
	=
	\gamma^d_1 + \frac{\tanh(7(t-\tau))}{2}(\gamma^d_2-\gamma^d_1)
	,
\end{equation*}
\ie, the rates transition smoothly from some initial rate ($\beta_1$ and $\gamma^d_1$) to some final rate ($\beta_2$ and $\gamma^d_2$) around time $\tau > 0$.

In the following, we fix $\tau = 2.1$ and an overall time interval of $[0,4]$.
The time interval $[1,3]$ is split into 4~disjoint subintervals, $\{(a_i,a_{i+1})\}_{i=1}^{4}$ (with $a_i = 1 + 0.5 \, (i-1)$) and the goal of the optimal design problem is to choose four times $e_i \in (a_i,a_{i+1})$, one in each interval, at which to measure the number of infected and deceased individuals for optimal inference of the 6~rates $\bm = [\beta_1, \alpha, \gamma^r, \gamma^d_1, \beta_2, \gamma^d_2]$.
We consider two ways to find the four optimal times $e_1,\ldots,e_4$:
\begin{enumerate*}[label=\ensuremath{(\roman*)}]
	\item \label[experiment]{item:SEIR:1}
		choosing them one at a time using a greedy sequential procedure (yielding the optimal design vector $\be_{\textup{SOED}-1}^*$), and

	\item \label[experiment]{item:SEIR:2}
		planning two subsequent observation times simultaneously using a greedy sequential procedure (yielding the optimal design vector $\be_{\textup{SOED}-2}^*$).
\end{enumerate*}
To set up the Bayesian inverse problem, a uniform prior on $[0,1]$ is assigned to each unknown rate, \ie, $\pi_\bm = \mathbbm{1}_{[0,1]^6}$.
The measurement noise at each observation time is assumed to be uncorrelated and Gaussian with zero mean and standard deviation $\sigma_I = 2$ for the measured number of infected individuals and $\sigma_{R^d} = 1$ for the measured number of deceased individuals.

\begin{figure}
	\centering
	\includegraphics[width = 0.85\textwidth]{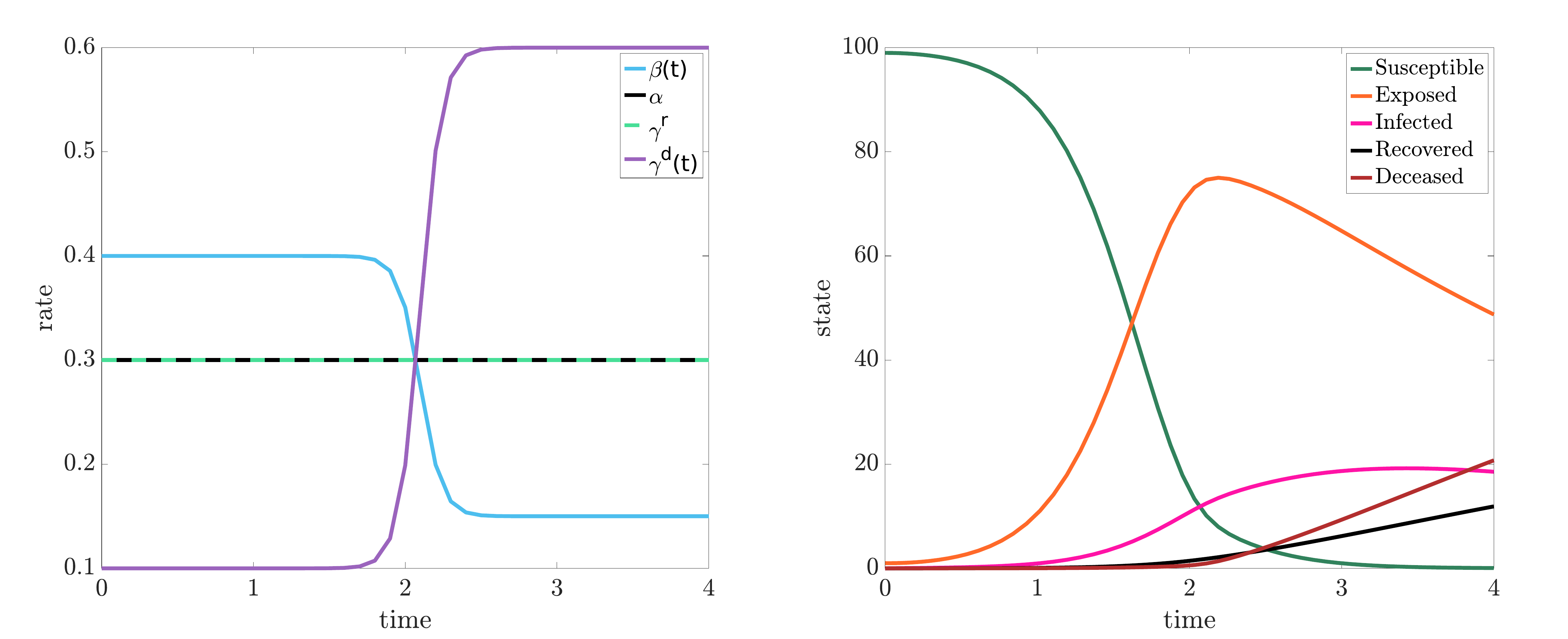}
	\caption{The true rates used for synthesizing data and the corresponding solution to the system~\eqref{eq:SEIR} are visualized on the left and right, respectively.}
	\label{figure:SEIR_visualization}
\end{figure}

We choose a fixed \enquote{true} parameter $\bm_{\textup{true}} = [0.4, 0.3, 0.3, 0.1, 0.15, 0.6]^\transp$ in the following experiments and focus on finding designs maximizing the EIG, \ie, the D-optimality criterion.
For all experiments, the dynamics are simulated in \matlab using \texttt{ode45} to solve the system of equations~\eqref{eq:SEIR}, while \matlab's \texttt{fmincon} function is used to maximize the objective function.
Simple bound constraints ensure that $e_i$ remains inside the corresponding time interval.
A visualization of the true rates as well as the corresponding solution to the SEIR model~\eqref{eq:SEIR} is provided in~\cref{figure:SEIR_visualization}

We begin with \cref{item:SEIR:2}.
Using~\cref{algorithm:OED} with $N = \num{10000}$ samples to approximate the EIG, we solve the greedy sequential design problem and plan the four optimal observation times two at a time, leading to the final optimal design vector~$\be_{\textup{SOED-2}}^*$.
Construction of the DIRT approximation to the two joint densities for the $4 + 2 + 6 = 12$-dimensional random variables $\be, \bd, \bm$ as well as the intermediate posterior density $\pi_{\bm \given e_{1:2}^*,d_{1:2}^*}$ required \num{350650} evaluations of the corresponding unnormalized densities, and thus \num{350650} solves of the SEIR ODE system, where each evaluation takes approximately \SI{3.5e-3}{\second} on average.

We also find the optimal allocation of testing times in a purely sequential fashion ($\be_{\textup{SOED}-1}^*$) for a fixed set of inference parameters using~\cref{algorithm:SOED_EIG}, \ie, \cref{item:SEIR:1}.
Again, $N = \num{10000}$ samples are used to approximate the incremental EIG in each stage of the SOED procedure.
The total procedure for finding four optimal observation times using the greedy sequential fashion required \num{147987} solves of the SEIR ODE system.
Note that if NMC were used to estimate the EIG with \num{10000} samples per evaluation, \num{15} EIG evaluations in the first SOED stage would require more solves of the SEIR system than our approach.
Furthermore, it is not straightforward to adjust the NMC estimator for the incremental EIG in an efficient way due to the need for posterior samples.

\begin{figure}[p]
	\centering
	\begin{tikzpicture}[rotate = 90, transform shape]
		\node [anchor = south west, inner sep = 0] (img) at (0,0) {\includegraphics[width = 1.1\linewidth]{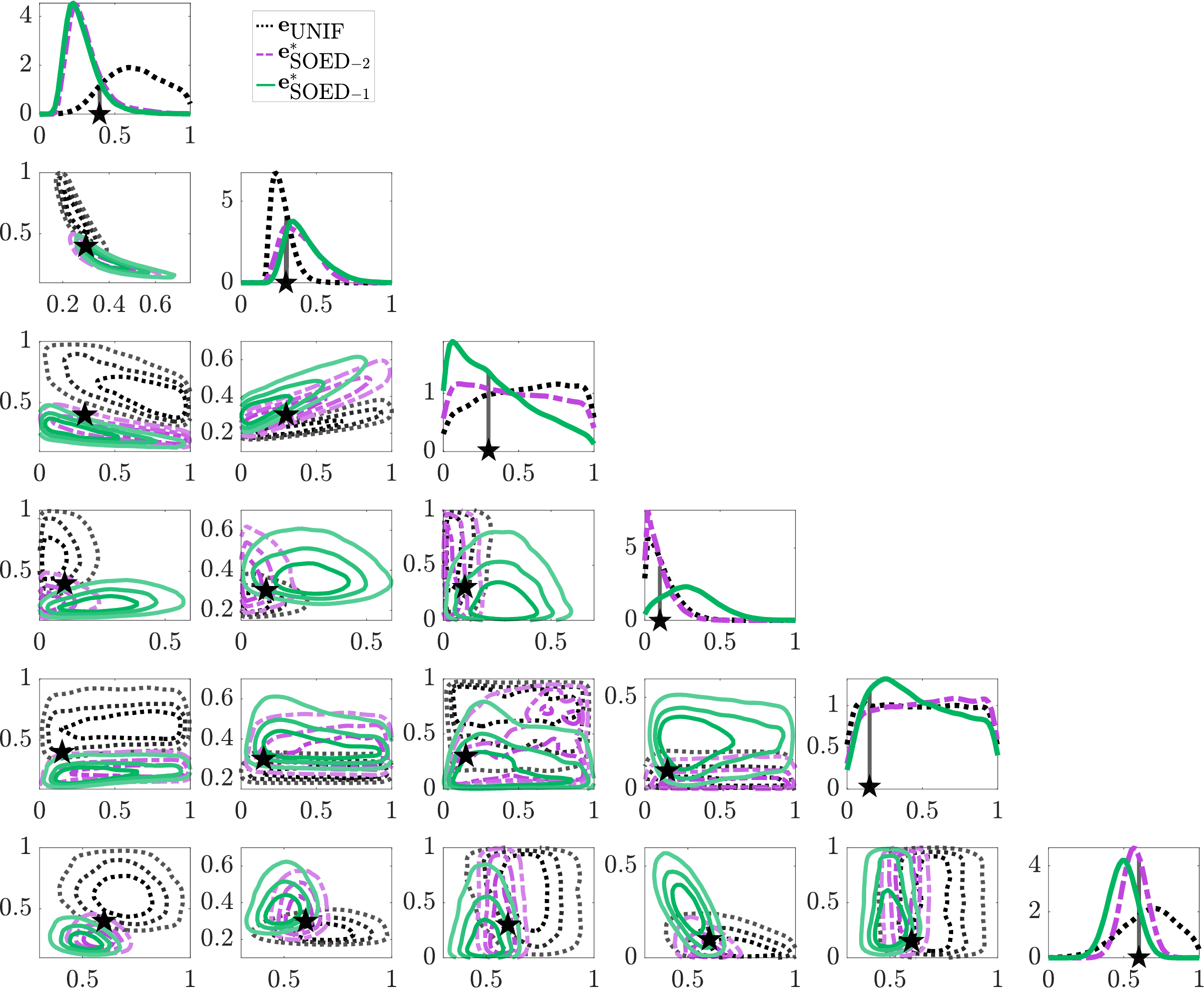}};
		\begin{scope}[x = {(img.south east)}, y = {(img.north west)}]
			\node [anchor = south] at (0.10,1.00) {$\beta_1$};
			\node [anchor = south] at (0.26,0.83) {$\alpha$};
			\node [anchor = south] at (0.43,0.66) {$\gamma^r$};
			\node [anchor = south] at (0.60,0.49) {$\gamma^d_1$};
			\node [anchor = south] at (0.77,0.32) {$\beta_2$};
			\node [anchor = south] at (0.93,0.15) {$\gamma^d_2$};
		\end{scope}
	\end{tikzpicture}
	\caption{Visualization of the marginals of the posterior densities obtained using different observation times for the SEIR model problem in~\cref{subsection:SEIR_example}.
		The posterior depicted using the black dotted line is obtained using the uniform observation times $\be_{\textup{UNIF}} = [1, 1.5, 2.0, 2.5]$.
		The purple dashed line is used to visualize the posterior obtained using the two-at-a-time greedy sequential designs $\be_{\textup{SOED}-2}^*$ and the solid green line corresponds to the posterior obtained using the one-at-a-time sequential optimal times $\be_{\textup{SOED}-1}^*$.
	In all cases, the data was synthesized using $\bm_{\textup{true}} = [0.4, 0.3, 0.3, 0.1, 0.15, 0.6]^\transp$, visualized with a black star and perturbed by noise.}
	\label{figure:SEIR_marginals}
\end{figure}
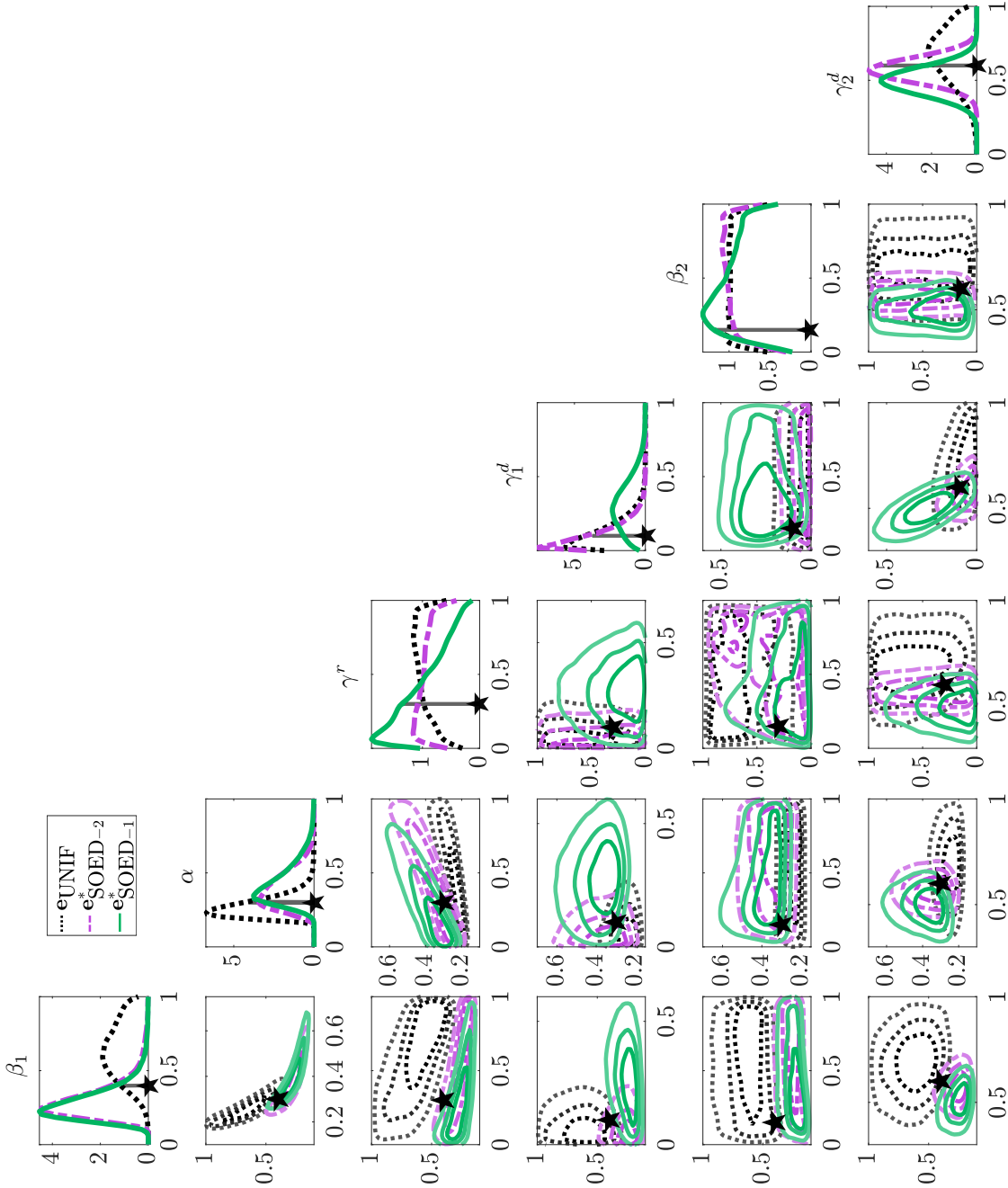

\Cref{figure:SEIR_marginals} provides a visual comparison of the posterior densities using data synthesized at the two \enquote{optimal} designs $\be_{\textup{SOED}-1}^*$ and $\be_{\textup{SOED}-2}^*$.
They are also compared to the posterior density resulting from data measured uniformly at the beginning of each time interval, $\be_{\textup{UNIF}} = [a_1,a_2,a_3,a_4]$.
For each design choice, the resulting posterior density was estimated using the \namedref{algorithm:DIRT}{DIRT Algorithm} such that $\distH{\widehat{\pi}_{\bm \given \be^*,\bd^*}}{\pi_{\bm \given \be^*,\bd^*}}$ is around $10^{-2}$.
Both optimal design choices lead to more concentrated posterior densities than $\be_{\textup{UNIF}}$ over most of the parameters, however $\be_{\textup{UNIF}}$ outperforms both optimal designs in learning the infection rate~$\alpha$.
The overall superiority of $\be_{\textup{SOED}-1}^*$ and $\be_{\textup{SOED}-2}^*$ is also supported by comparing the KL divergence from posterior to prior.
Using \num{100000} samples to approximate the KL divergence, we have:
\begin{align*}
	\distKL{\pi_{\bm \given \be_{\textup{UNIF}}, \bd_{\textup{UNIF}}}}{\pi_\bm}
	&
	\approx
	5.1
	,
	\\
	\distKL{\pi_{\bm \given \be_{\textup{SOED}-1}^*, \bd_{\textup{SOED}-1}}}{\pi_\bm}
	&
	\approx
	6.3
	,
	\\
	\text{and}
	\quad
	\distKL{\pi_{\bm \given \be_{\textup{SOED}-2}^*, \bd_{\textup{SOED}-2}}}{\pi_\bm}
	&
	\approx
	6.8
	.
\end{align*}
While the simultaneous planning of two experiments at a time in the greedy SOED procedure requires more solves of the SEIR system, it leads to a much better estimate of the initial rate of mortality ($\gamma_1^d$) and a slightly better estimate of the final rate of mortality ($\gamma_2^d$), in alignment with the larger value of the KL divergence.
\begin{remark}[On the lack of information about the final rate of exposure $\beta_2$]
	We note that all the designs struggle with learning the final rate of exposure $\beta_2$, but this is a consequence of the transition time lying quite late in the simulation interval.
	Once the policies are implemented, almost all susceptible individuals are already exposed, hence the difficulty.
	We have also performed this experiment with a smaller value of $\tau$, in which case $\beta_2$ is more accurately estimated at the expense of the precision of some of the initial rates, \eg, the initial mortality rate $\gamma_d$.
	\label{remark:unlearnable_rates}
\end{remark}

\begin{figure}[htp]
	\centering
	\includegraphics[width = 0.6\linewidth]{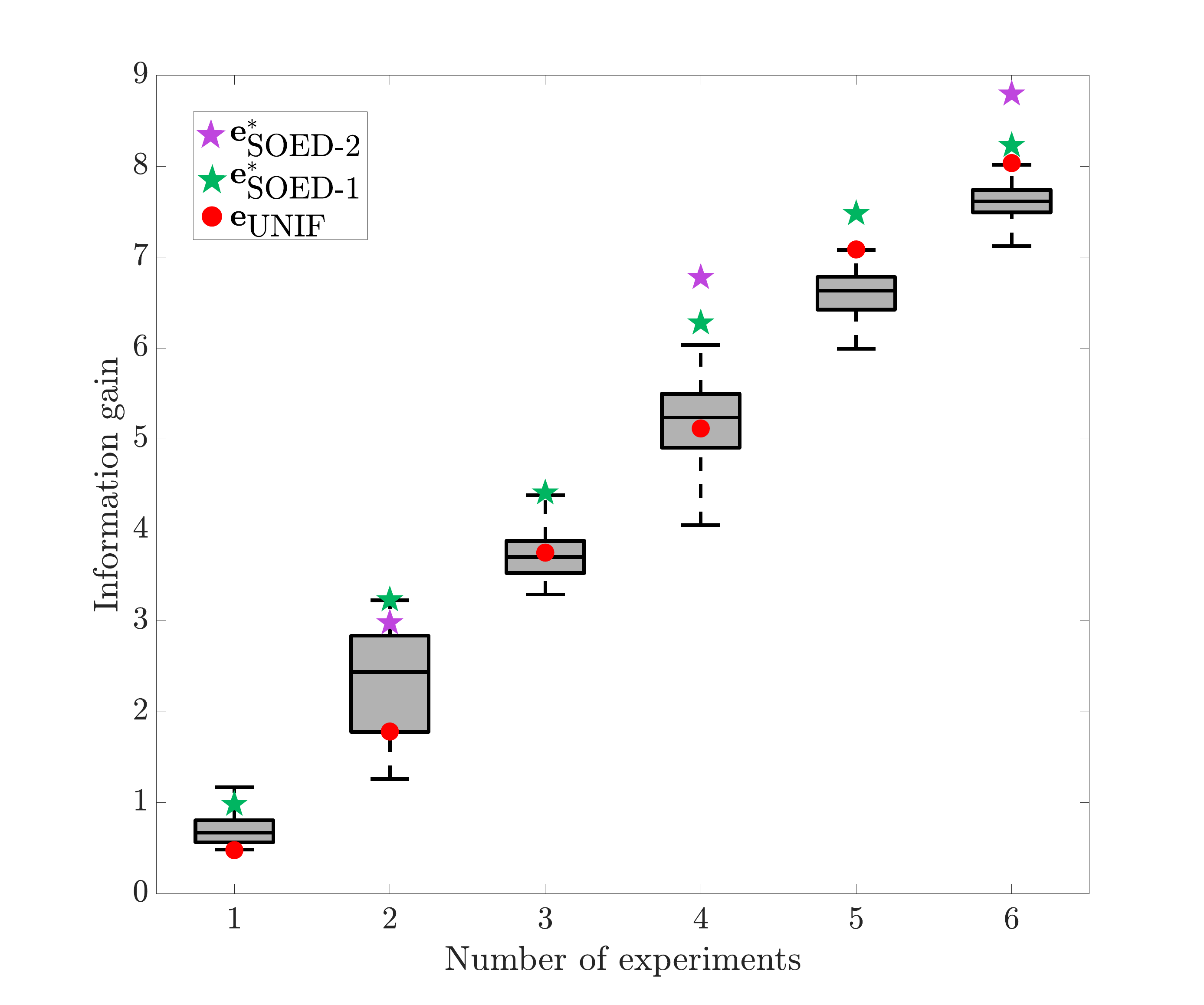}
	\caption{Boxplot comparing the effectiveness of the optimal observation times obtained for the SEIR example defined in~\cref{subsection:SEIR_example} using the two-at-a-time and one-at-a-time procedures described in~\cref{subsection:SEIR_example} (visualized using purple and green stars, resp.), randomly chosen observation times (visualized in black), and the uniform design $\be_{\textup{UNIF}}$ (red circles).
		To obtain the results, data was synthesized using $\bm_{\textup{true}}$ for the optimal designs ($\be_{\textup{SOED}-1}^*$ and $\be_{\textup{SOED}-2}^*$) and for $100$ randomly chosen observation times.
	For each set of data and design, the information gain or $\distKL{\pi_{\bm \given \be, \bd}}{\pi_\bm}$ was approximated using \num{100000} samples from a DIRT approximation to the posterior density $\pi_{\bm \given \be, \bd}$.}
	\label{figure:SEIR_IG}
\end{figure}

In~\cref{figure:SEIR_IG}, we compare the performance of the optimal designs, randomly chosen designs, and $\be_{\textup{UNIF}}$.
For this comparison, we extended the observation times to include two additional intervals.
The optimal observation times outperform the as-soon-as-possible uniform observation times and the randomly chosen designs, and the gap greatly increases after the first experiment, since our observation times can be better geared to~$\bm_{\textup{true}}$ due to the feedback loop.
It slowly tapers off after four experiments once information begins to saturate.
Interestingly, $\be_{\textup{UNIF}}$ starts off as the worst design choice but after 6~experiments performs almost as well as the one-at-a-time greedy sequential design $\be_{\textup{SOED}-1}^*$.

\subsection{Optimal Dirichlet data for permeability field inversion}
\label{subsection:Poisson_example}

Here, we consider the elliptic PDE
\begin{equation}
	- \div \paren[auto](){\kappa \nabla u}
	=
	0
	\quad
	\text{on }
	\Omega
	=
	(0,1)^2
	,
	\label{eq:elliptic}
\end{equation}
which is often used in groundwater modeling.
In this example, the inverse problem consists of estimating the spatially-dependent diffusivity field~$\kappa$, given measurements of the pressure~$u$ at some pre-determined locations $(x_i,y_i) \in \Omega$.
To ensure~$\kappa$ is nonnegative, we impose a Gaussian prior on the log diffusivity, $m = \log(\kappa) \sim \normal{0}{\Cpr}$, with covariance operator $\Cpr$ defined using a squared-exponential kernel
\begin{equation*}
	c(x,z)
	=
	\sigma_v^2 \exp \paren[auto][]{ \frac{- \norm{x-z}^2}{2 \, \ell^2}}
	\quad
	\text{for }
	x,z \in \Omega
	,
\end{equation*}
with $\sigma_v = 1$ and $\ell^2 = 0.1$.
Employing a truncated Karhunen-Loève expansion of the unknown diffusivity field yields the approximation
\begin{equation*}
	m(x, \bm)
	\approx
	\sum_{i=1}^\Nm m_i \sqrt{\lambda_i} \, \phi_i(x)
	,
\end{equation*}
where $\lambda_i$ and $\phi_i(x)$ denote the the $i$-th largest eigenvalue and eigenfunction of~$\Cpr$, respectively, and the unknown coefficients $m_i \sim \normal{0}{1}$.
The Karhunen-Loève expansion is truncated after $\Nm = 16$~modes, resulting in an approximation that captures $99$~percent of the weight of~$\Cpr$.

\begin{figure}[htp]
	\centering
	\begin{tikzpicture}
		\node [anchor = south west, inner sep = 0] (img) at (0,0) {\includegraphics[width = \textwidth]{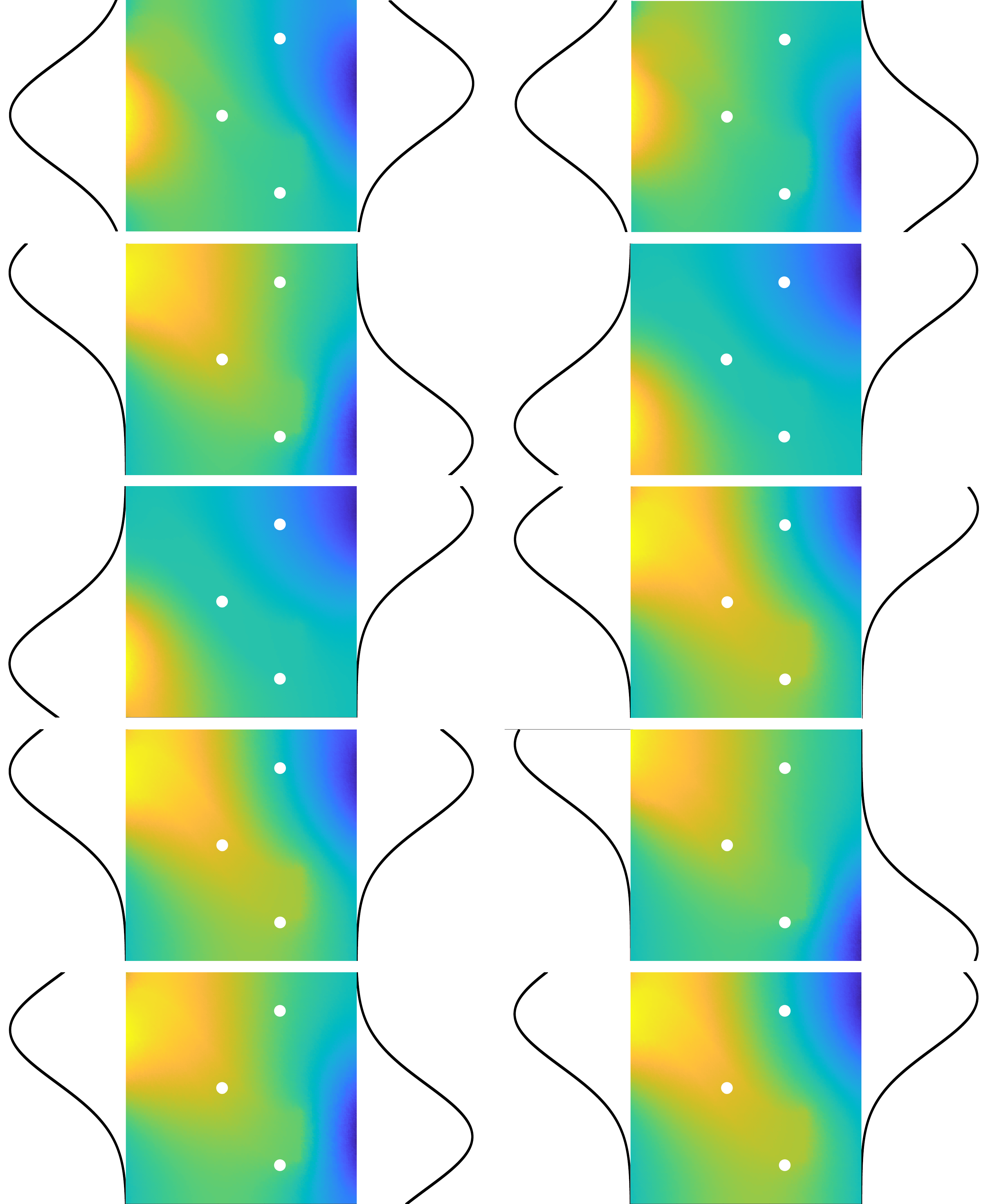}};
		\begin{scope}[x = {(img.south east)}, y = {(img.north west)}]
			\node [anchor = south] at (0.25,1.0) {$\be_A^*$};
			\node [anchor = south] at (0.75,1.0) {$\be_D^*$};
		\end{scope}
	\end{tikzpicture}
	\caption{Visualization of optimal boundary conditions chosen using the greedy sequential procedure with the A-optimality criterion (left column) and D-optimality criterion (right column). The boundary conditions are plotted on the left and right boundaries in each figure and the corresponding state~$u$ is visualized for the true log diffusivity field.}
	\label{figure:boundary-conditions}
\end{figure}

\begin{figure}[htp]
	\centering
	\begin{tikzpicture}
		\node [anchor = south west, inner sep = 0] (img) at (0,0) {\includegraphics[width = \textwidth]{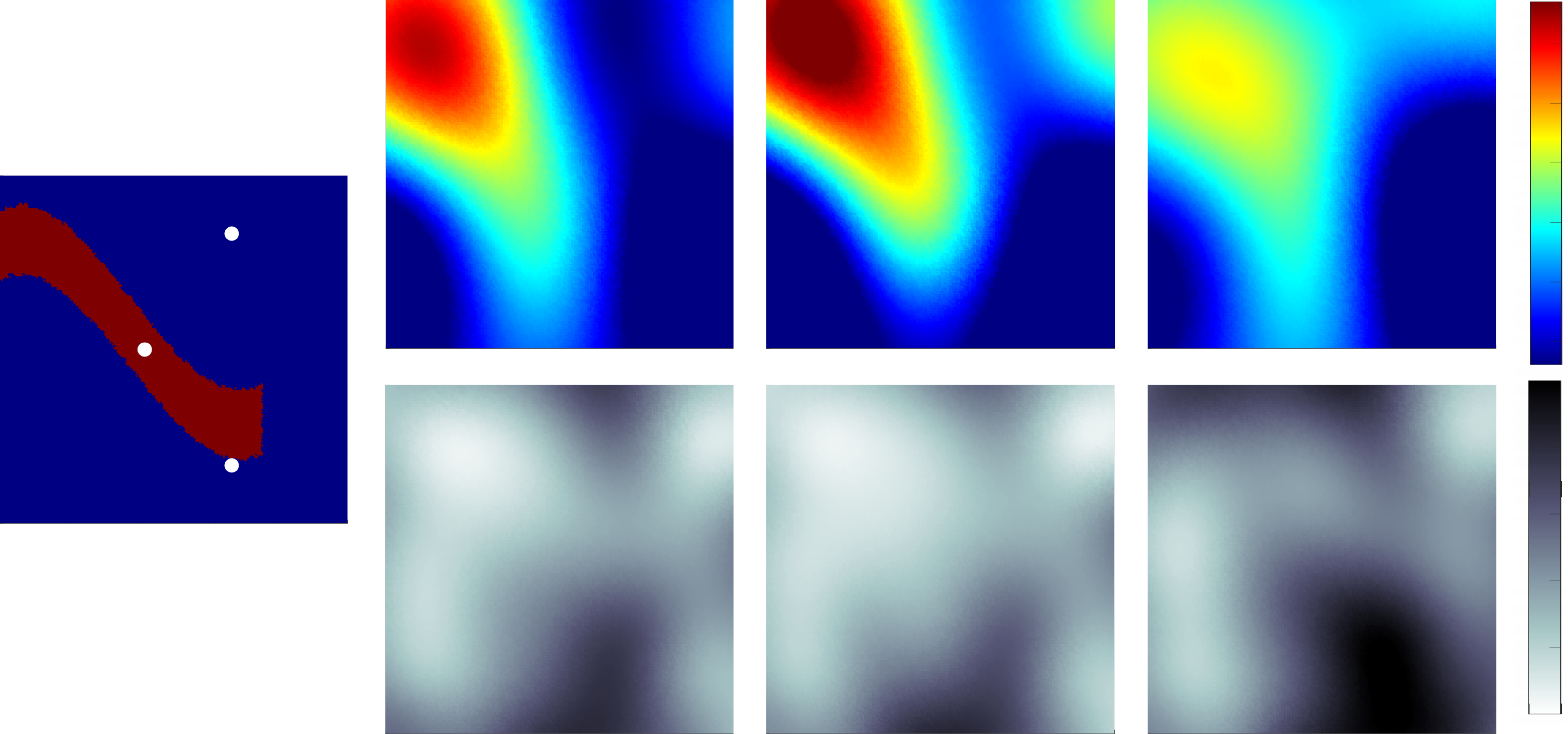}};
		\begin{scope}[x = {(img.south east)}, y = {(img.north west)}]
			\node at (0.04,0.32) {\textcolor{white}{$m_\textup{true}$}};
			\node [anchor = south] at (0.37,1.0) {$\be_D^*$};
			\node [anchor = south] at (0.61,1.0) {$\be_A^*$};
			\node [anchor = south] at (0.85,1.0) {$\be_{\textup{rand}}$};
			\node [anchor = west] (topmin) at (1.0,0.52) {\small $-0.5$};
			\node [anchor = west] (topmax) at (1.0,1.00) {\small $\mrep[l]{2.5}{-0.5}$};
			\node [anchor = center] at ($(topmin)!0.5!(topmax)$) {\small $\mrep[l]{1.0}{-0.5}$};
			\node [anchor = west] at (1.0, 0.03) {\small $0.3$};
			\node [anchor = west] at (1.0, 0.47) {\small $0.8$};
		\end{scope}
	\end{tikzpicture}
	\caption{On the far left, the true log-diffusivity field used for synthesizing the pressure data at the three locations (visualized as white dots) is given.
		The top row in the $2 \times 3$ array of images visualizes the posterior mean obtained using five experiments with boundary conditions chosen using the greedy SOED procedure with the D-optimality criterion ($\be_D^*$) and A-optimality criterion ($\be_A^*$), and a randomly chosen design ($\be_{\textup{rand}}$).
		The bottom row visualizes the corresponding posterior pointwise variance for each design choice.
	The posterior means and variances were approximated using \num{100000} samples from the corresponding approximate posterior density.}
	\label{figure:true_kappa}
\end{figure}

In this example, we consider an intrusive design, \ie, we choose Dirichlet data to impose at the left and right boundary, which we parametrize as
\begin{subequations}
	\begin{align}
		u(x = 0,y)
		&
		=
		\exp \paren[Big](){-\frac{1}{2\sigma_w} \paren[auto](){y-e_1}^2}
		,
		\\
		u(x=1,y)
		&
		=
		- \exp \paren[Big](){-\frac{1}{2\sigma_w} \paren[auto](){y-e_2}^2}
		.
	\end{align}
\end{subequations}
Homogeneous Neumann data is fixed at the top and bottom boundaries.
Thus, the design $\be = [e_1,e_2]$ enters into the parameter-to-state map directly through the boundary condition.
The effect of different designs on the state can be seen in \cref{figure:boundary-conditions}.
After each experiment is performed with some prescribed boundary conditions, $u$~is measured at three locations as visualized in~\cref{figure:true_kappa}.

Using~\cref{algorithm:SOED_EIG}, boundary conditions are chosen for five experiments in a sequential fashion.
The log-diffusivity field used to synthesize the data in each experiment is visualized in~\cref{figure:true_kappa}.
At each stage, the data is synthesized using the Fast Forward and Inverse problems solver (\texttt{FastFins}) package~\cite{Cui:fastfins}.
Specifically, the governing PDE~\eqref{eq:elliptic} is solved using the finite element (FE) method with second-order Lagrange elements on a mesh of size $h = \frac{1}{64}$ in each coordinate direction.
Applying the resulting discretized PTO map takes approximately \SI{0.17}{\second}.
To speed up the construction of the FTT approximation to the joint density $\pi_{\be, \bd, \bm \given \bI_k}$ in each stage, the design-dependent parameter-to-state map is replaced with a surrogate built using the discrete empirical interpolation method~\cite{ChaturantabutSorensen:2010:1}.
The surrogate is constructed using the \texttt{FastFins} package with \num{1000} solves of the full-order elliptic PDE, and has relative approximation error on the order of $10^{-3}$.
One solve with the reduced-order model is approximately $60$~times faster than one solve of the FE model.

As in the disease modeling example outlined in the previous section, in each stage of the SOED procedure, the bridging densities are obtained by tempering the likelihood.
For this example, we construct a sequence of greedy optimal designs using both A- and D-optimality criteria.
In the case of D-optimality, $N = \num{1000}$ samples are used to approximate the expected information gain.
To approximate the A-optimal sequential designs, $N = \num{500}$ Quasi-Monte Carlo samples are used for the outer expectation with respect to the evidence, and $M = \num{1000}$ Monte Carlo samples are used for the inner expectation.
\Cref{table:cost} presents the computational cost (given in terms of FE solves) for constructing DIRT approximations to the joint density $\pi_{\be_k, \bd_k, \bm \given \bI_{k-1}}$ and to the posterior $\pi_{\bm \given \bI_k}$ in each SOED stage, $k = 1, \ldots, 5$, as well as the cost for evaluating $\PsiA(\be_k)$ and $\PsiD(\be_k)$ once for any choice of design $\be_k \in \eSpace_k$ in the optimization routine.
For this particular problem, the A-optimal utility function exhibits smaller variations for different designs than the D-optimal one, so higher accuracy in the FTT surrogates to the joint densities are required (particularly in the later stages) to capture the correct \enquote{valleys} and \enquote{peaks}.

\begin{table}[htp]
	\centering
	\begin{tabular}{|c|c|c|c|}
		\hline
		& \multicolumn{1}{l|}{\textbf{Stage ($k$)}} & \textbf{A-optimal} & \textbf{D-optimal} \\ \hline
		\multirow{5}{*}{\textbf{$\pi_{\be_k, \bd_k, \bm \given \bI_{k-1}}$}} & 1                                         & 4568               & 4339               \\ \cline{2-4}
		& 2                                         & 5850               & 4403               \\ \cline{2-4}
		& 3                                         & 5403               & 3058               \\ \cline{2-4}
		& 4                                         & 4568               & 1747               \\ \cline{2-4}
		& 5                                         & 3945               & 1898               \\ \hline
		\multirow{5}{*}{\textbf{$\widehat{\Psi}_X^k(\be_k)$}}                & 1                                         & 139                & 25                 \\ \cline{2-4}
		& 2                                         & 87                 & 24                 \\ \cline{2-4}
		& 3                                         & 116                & 41                 \\ \cline{2-4}
		& 4                                         & 139                & 50                 \\ \cline{2-4}
		& 5                                         & 156                & 59                 \\ \hline
		\multirow{5}{*}{\textbf{$\pi_{\bm \given \bI_k}$}}                   & 1                                         & 632                & 595                \\ \cline{2-4}
		& 2                                         & 409                & 518                \\ \cline{2-4}
		& 3                                         & 494                & 419                \\ \cline{2-4}
		& 4                                         & 632                & 258                \\ \cline{2-4}
		& 5                                         & 428                & 309                \\ \hline
	\end{tabular}
	\caption{Computational cost, presented in terms of the number of FE solves of the underlying PDE, for constructing the DIRT approximations to $\pi_{\be_k, \bd_k, \bm \given \bI_{k-1}}$ and $\pi_{\bm \given \bI_k^*}$, as well as the evaluation of the incremental optimality criterion $\widehat{\Psi}_X^k$, in each SOED stage~$k$.
		Note that the A-optimality criterion $\EhatPsiA^k$ was evaluated in parallel using $50$~workers, whereas the D-optimality criterion was evaluated in serial.
	However, the D-optimality criterion could also be easily parallelized.}
	\label{table:cost}
\end{table}

\begin{figure}[htp]
	\centering
	\includegraphics[width = \textwidth]{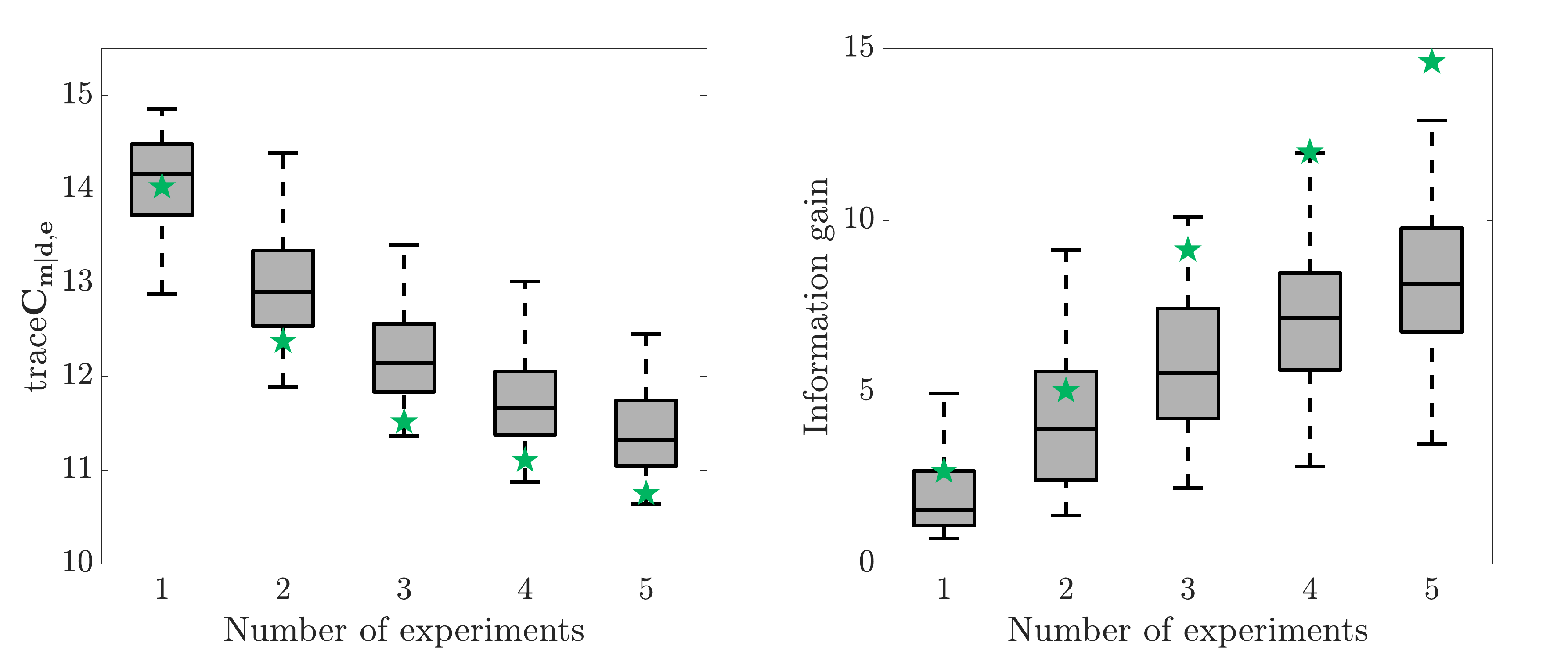}
	\caption{On the left, $\trace \paren[auto](){\bC_{\bm \given \bI_k}}$, and on the right, $\distKL{\pi_{\bm \given \bI_k}}{\pi_\bm}$, from conducting $k = 1, \ldots, 5$ experiments.
		The green stars are obtained by conducting the experiments with boundary conditions chosen using the SOED procedure with the A- and D-optimality criterion.
		Note that the A-optimal design should maximize $-\trace \paren[auto](){\bC_{\bm \given \bI_k}}$, hence minimize $\trace \paren[auto](){\bC_{\bm \given \bI_k}}$.
		We plot the latter, since the A-optimal design is often interpreted as the one minimizing the average posterior pointwise variance.
		The box plot is obtained by using 100~randomly chosen boundary conditions for all five experiments.
	In all cases, data was synthesized using the true diffusivity field visualized in~\cref{figure:true_kappa}.}
	\label{figure:Poisson_IG}
\end{figure}

\Cref{figure:true_kappa} compares the means and the pointwise variances of the posterior distribution $\pi_{\bm \given \be_{1:5}, \bd_{1:5}}$ using synthesized data from five experiments conducted with boundary conditions chosen randomly $(i)$, as well as using the SOED procedure with both utility functions, $\psi^k_A$ $(ii)$ and $\psi^k_D$ $(iii)$.
As made evident by both visualizations, the optimal designs perform much better at recovering the diffusivity channel.
This observation is further strengthened in~\cref{figure:Poisson_IG}, where both optimality criteria, $\psiA^k$ and $\psiD^k$, are evaluated at the A- and D-optimal sequence of boundary conditions as well as at randomly chosen boundary conditions with data synthesized using the true log-diffusivity field.

While other methods like the nested Monte Carlo approach (\eg, see~\cite{HuanMarzouk:2013:1}) may need fewer samples in batch OED settings, they are generally not applicable to SOED without significant modifications.
In contrast, our transport map approach adapts well to sequential settings.
As shown in~\cref{table:cost}, computational costs tend to decrease across experimental stages due to the preconditioning approach described in~\cref{subsection:soed_preconditioners}.
Moreover, after conducting experiments, the KR rearrangement provides an approximation to the posterior that can either be used directly or to enhance MCMC sampling efficiency.

\section*{Acknowledgements}

This work has been partially funded by the Carl Zeiss-Stiftung through the project “Model-Based AI: Physical Models and Deep Learning for Imaging and Cancer Treatment”.
The authors would like to thank Tiangang Cui and Sergey Dolgov for many helpful discussions and access to the developmental version of the DIRT \matlab package.
The authors thank the reviewers for helpful comments and suggestions.

\appendix

\printbibliography

\end{document}